\documentclass[aps,prc,nofootinbib,twocolumn,superscriptaddress,showpacs,
floatfix]{revtex4-1}
\usepackage{mathrsfs}
\usepackage{latexsym}
\usepackage{amsmath}
\usepackage{amssymb}
\usepackage{graphicx}
\usepackage{epsfig}
\usepackage[usenames]{color}
\usepackage[english]{babel}
\newcommand \beq{\begin{eqnarray}}
\newcommand \eeq{\end{eqnarray}} 
\def\x{{\boldsymbol x}}
\def\p{{\boldsymbol p}}

\begin{document}

\title{Relativistic Boltzmann transport approach with Bose-Einstein statistics \\
and the onset 
of gluon condensation}

\author{F. Scardina}
\affiliation{Physics and Astronomy Dept., University of Catania, Via S. Sofia 64, I-95123
Catania, Italy}
\affiliation{INFN-Laboratori Nazionali del Sud, Via S. Sofia 62, I-95123
Catania, Italy}

\author{D. Perricone}
\affiliation{Physics and Astronomy Dept., University of Catania, Via S. Sofia 64, I-95123
Catania, Italy}
\affiliation{INFN-Laboratori Nazionali del Sud, Via S. Sofia 62, I-95123
Catania, Italy}

\author{S.Plumari}
\affiliation{Physics and Astronomy Dept., University of Catania, Via S. Sofia 64, I-95123
Catania, Italy}
\affiliation{INFN-Laboratori Nazionali del Sud, Via S. Sofia 62, I-95123
Catania, Italy}

\author{M. Ruggieri}
\affiliation{Physics and Astronomy Dept., University of Catania, Via S. Sofia 64, I-95123
Catania, Italy}

\author{V. Greco}
\affiliation{Physics and Astronomy Dept., University of Catania, Via S. Sofia 64, I-95123
Catania, Italy}
\affiliation{INFN-Laboratori Nazionali del Sud, Via S. Sofia 62, I-95123
Catania, Italy}


\begin{abstract}
We study the evolution of a gluon system under conditions of density and temperature similar 
to those explored in the early stage of ultra-relativistic heavy-ion collisions. 
We first describe the implementation of Relativistic Boltzmann-Nordheim (RBN) transport approach 
that includes in the collision integral the quantum effects of Bose-Einstein Statistics. 
Then, we describe the evolution of a spatially uniform gluon system in a box under elastic collisions solving
the RBN for various initial conditions. We discuss the critical phase-space density that leads to the onset
of a Bose-Einstein condensate (BEC) and the time scale for this process to occur.
In particular, thanks to the fact that RBN allows to relax the small angle approximation, we study the effect
at both small and large screening mass $ m_{D} $. For small $ m_{D}\ll T $ we see that our solution
of RBN is in agreement with the recent extensive studies within a Fokker-Planck scheme in small angle approximation.
For the same total cross section but with large $ m_{D}\simeq 2\, T $ (large angle scatterings), we see a significant time speed-up of the onset of BEC respect to small $m_{D}\ll T$. This further strengthen the possibility
that at least a transient BEC is formed in the early stage of ultra-relativistic heavy-ion collisions.

\end{abstract}


\maketitle

\section{Introduction}

The experiments of ultra relativistic heavy ion collisions performed at RHIC and LHC  have given clear indication that a hot and dense
strongly interacting quark and gluon plasma (QGP) can be created in laboratory \cite{Adams:2005dq,Adcox:2004mh,Aamodt:2010pa,Fries:2008hs,Jacak:2012dx}. 
The dynamical behavior of such a state of matter and in particular its strong anisotropic collective expansion
can be described by means of few parameters by viscous hydrodynamics \citep{Luzum:2008cw,Hirano:2009ah,Alver:2010dn,Song:2011hk,Niemi:2011ix} with the assumption of an early thermalization time $\tau \sim 0.5-1\ \mathrm{fm/c}$. 
However it has been argued recently that the initial gluonic systems created in the very early stage, before thermalization, is so dense that its quantum Bose-Einstein nature plays an important role eventually driving the system toward at least a transient Bose-Einstein Condesate (BEC) \cite{Blaizot:2011xf,Blaizot:2012qd,Blaizot:2013lga,Berges:2012us,Dusling:2010rm,Huang:2013lia}.
Such a picture is in direct agreement with  a Color Glass Condensate (CGC) theory.
\cite{McLerran:1993ni,McLerran:1993ka,McLerran:1994vd,Gelis:2010nm,Kharzeev:2001gp,Kharzeev:2004if}. 
In fact,
the very high density of the gluon distribution functions at low $x$ of 
the incoming nuclei triggers a saturation of the initial momentum distribution of the matter below a saturation scale $Q_s$ having also
an occupation number $f \sim 1/\alpha_{s}$. 
In this framework it is expected that the gluon density in the initial stage is large enough so that the system contains more gluons than can be accommodated by a Bose - Einstein (BE)
equilibrium distribution as has suggested initially in \cite{Blaizot:2011xf}. 
More specifically, the dimensionless quantity $n\epsilon^{-3/4}$, where $n$ is the gluon density and 
$\epsilon$ the energy density, exceeds the value for a system of gluons at thermodynamic equilibrium. 
If this is the case and 
if the mechanism of approaching equilibrium is dominated by processes which conserve the total number of particles then a Bose condensate will develop. 
The impact of the quantum nature of bosons have been recently discussed also for light ion production
in heavy-ion collisions at intermediate energy \cite{Zheng:2011ke,Giuliani:2013kna}.

The evolution of a gluon system towards the condensate have been thoroughly studied in 
\cite{Blaizot:2013lga} for a static medium using the Fokker Planck approach in which the effect of the Bose-Einstein statistics have been taken into account. 
More recently also the impact of finite quark density has been discussed in \cite{Blaizot:2014jna}.
The Fokker-Planck approach is an approximation of the Boltzmann transport equation that is strictly valid in the small angle scattering limit. The Fokker-Plank equation is easier to solve with respect
to the Boltzmann equation but has the advantage to supply a more transparent description of the
underlying dynamics. Nonetheless, as pointed out in \cite{Das:2013kea}, it may not encase all the dynamics of the collisions when the system does not evolve only through soft scatterings. 
Several studies of in-medium dynamics suggest  the presence of large Debye screening mass
$m_{D}\simeq g(T) \, T\sim 1 \, \rm GeV$ at temperature typical of uRHIC's which would 
determine scatterings with $q^2\gg T^2$.
For this reason we study the evolution of a gluon system along similar line as in \citep{Blaizot:2013lga,Huang:2013lia} but solving numerically the full relativistic Boltzmann-like equation.

The Relativistic Boltzmann approach has been developed  to study the 
evolution of the QGP in ultra-relativistic heavy 
ion collision (uRHIC's) and in particular the elliptic flow estimating the shear viscosity to entropy density to be about 
 $\eta/s \simeq 0.1-0.2$ \cite{Ferini:2008he,Xu:2008av,Xu:2007jv,Plumari:2011re,Ruggieri:2013ova,Ruggieri:2013bda}, in agreement with viscous hydrodynamics approach.
The Boltzmann equation describing the evolution of $f$ can be compactly written as:
\begin{equation}
 p^{\mu} \partial_{\mu}f(x,p)= {\cal C}[f](x,p)
\label{B_E}
\end{equation}
where ${\cal{C}}[f](x,p,t)$ is the collision integral. The one-body distribution function in our
case can be written as:
\beq
\label{dis_funct}
    f(\x,\p)=\frac{(2\pi)^3}{d_A}\frac{dN}{d^3\x d^3\p},
\eeq
with $d_A={2(N_c^2-1)}$ corresponding to the  degrees of freedom for gluons.
To our knowledge it has always been neglected the bosonic nature of particles when applying the 
Boltzmann equation to uRHIC, which instead under certain conditions can strongly
determine the phase space evolution. This choice has been driven by the fact that during 
the evolution of the system the density is small enough to make such  quantum corrections negligible. 
However as mentioned above this assumption could not be valid in the very early times of the evolution of the matter that comes up after the collision. 
We will describe in this paper the implementation of numerical solutions of a Boltzmann-like transport equation
having as fixed point the BE distribution function. Similar approaches have been already developed in a non-relativistic regime to study ultra-cold atomic systems \cite{PhysRevA.66.033606,Pantel:2012pi}.

The article is organized as follows. In the next section we will discuss the Boltzmann collision integral including the Bose-Einstein statistics
and we will describe the simulations code giving some details of the numerical implementation of the collision integral. In
section III, we discuss the initial condition that we have used to study the evolution of the gluonic system toward a condensate, or toward a Bose-Einstein 
equilibrium distribution, if the density is not enough large. The section IV and V are devoted to the numerical
results. Section VI contains summary and conclusions.

\section{Collision integral with Bose-Einstein statistics}

\subsection{Numerical setup}
In this section we describe the numerical code we have implemented
to solve the kinetic
equation improved with respect to \citep{Lang1993391,Zhang:1998tj,Molnar:2001ux,Xu:2004mz,Ferini:2008he} to take into account the quantum statistics in the collision integral that for the Boltzmann statistics has the form:
\begin{eqnarray}
C[f]&=&\frac{1}{2E_1}\int \frac{d^3p_2}{2E_2(2\pi)^3}\frac{1}{\nu} \int\frac{d^3p_{1}^{\prime}}{2E_{1'}(2\pi)^3} \int\frac{d^3p_{2}^{\prime}}{2E_{2'}(2\pi)^3}\nonumber\\
&&\times \left[f(p_{1}^{\prime})f(p_{2}^{\prime}) - f(p_1) f(p_2)\right]|{\cal M}(p_1p_2\rightarrow p_1^\prime p_2^\prime)|^2 \nonumber \\
&&\times(2\pi)^4\delta^4(p_1 + p_2 - p_{1}^{\prime} - p_{2}^{\prime})~,
\label{CollisionInt0}
\end{eqnarray}
where ${\cal M}$ corresponds to the transition amplitude; $\nu$ is set to 2 if one considers identical particles, otherwise is set to 1.
In the above equation only the two body collision term has been considered.
The quantum Bose-Einstein statistics is achieved by the replacement 
\begin{eqnarray}
&&f(p_{1}^{\prime})f(p_{2}^{\prime}) - f(p_1) f(p_2)\rightarrow\nonumber\\  
&&f(p_{1}^{\prime})f(p_{2}^{\prime}) (1+f(p_1))(1+f(p_2)) - \nonumber\\
&& f(p_1) f(p_2)(1+f(p_{1}^{\prime}))(1+f(p_{2}^{\prime}))~
\label{one_plus_f}
\end{eqnarray}
in the kernel of Eq.~\eqref{CollisionInt0} that is often renown as the Boltzmann-Nordheim equation.

In the present work we consider a system in a static box made of gluons 
interacting via elastic two body collisions. 
The Boltzmann equation is solved numerically on a space-time grid as described in \cite{Ferini:2008he,Xu:2004mz,Lin:2004en,Scardina2013296}, and
we use the standard test particle method to sample the distributions functions.
We have used $10^4$ test particles per real one for a total of $10^6-10^7$
test particles. 
The solution of the transport equation is equivalent to solve the 
Hamilton equations of motion for the test particles: 
coordinate $\bm r$ of the test particle at time $t^+$
is related by that at time $t^-$ by
\begin{eqnarray}
\label{Hamiltons}
\boldsymbol{r}(t^+)&=&\boldsymbol{r}(t^-)+ \, \Delta t \,\frac{\boldsymbol{p}(t^-)}{E(t^-)}
\end{eqnarray}
being $\Delta t$ the numerical mesh time.
While the momenta of the test particles are changed because of the collisions according to the two body relativistic kinematics.
In order to compute the collision integral we use a stochastic method 
in which the collisions among test particles are determined by the collision probability 
that can be derived 
directly from the collision integral in Eq. \eqref{CollisionInt0}, as shown in appendix A; 
for Boltzmann statistics the collisional probability has the form
\begin{equation}
\label{p22}
P^{BE}_{22} = \frac{\Delta N_{coll}^{2\to 2}}{\Delta N_1 \Delta N_2} = 
v_{rel} \sigma_{22} \frac{\Delta t}{\Delta^3 x}\,.
\end{equation}
In the above equation $\Delta^3 x$ is the volume of the grid cells;
$\Delta t$ is the mesh time of the simulations, $\Delta N$ is the number of particles inside a cell and $v_{rel}=s/2E_1E_2$ denotes the relative velocity,
where $s$ is Mandelstam variable relative to particles pair.
In the stochastic approach, for each cell of the grid and at each time step, 
we evaluate the collision probability $P_{22}$ between all the possible pairs of particles
and we compare it with a random number between 0 and 1: 
if the extracted number is smaller than $P_{22}$ then the collision occurs and the code
evaluates the final momenta of the colliding particles according to the angular
 dependence of the scattering matrix elements $|{\cal M}|^2$. 
In order to reduce the computational time,
instead of evaluating probabilities of all the pairs of particles usually one proceeds as indicated in Refs.\cite{Xu:2004mz,Danielewicz:1991dh} 
choosing randomly ${\cal N}$ out of the possible doublets and amplifying the collision probability by a factor $k$ 
\begin{equation}
\label{prob_mod}
k=\frac{n(n-1)}{{2\cal N}} 
\end{equation}
where $n$ is the number of particles inside a cell. The choice of ${\cal N}$ is arbitrary, however a good compromise between a substantial reduction of the 
computational time and avoiding a probability larger than 1 is to fix
 ${\cal N}$ equal at least to the number $n$ of particles inside a cell.
 
A relation for the collisional probability 
can also be obtained for the case of BE statistics, namely
\begin{eqnarray}
P_{22}=\frac{\Delta t}{\Delta^3x} 
\int d\Omega \frac{d\sigma}{d\Omega}(1+f(p{^\prime}_1))(1+f(p{^\prime}_2)){v}_{rel}
\end{eqnarray}
The derivation of $P^{BE}_{22}$ is shown in Appendix A and in reference \cite{PhysRevA.66.033606}.
The difference between $P_{22}$ and $P^{BE}_{22}$ is due to the presence of the terms $(1+f(p_{1}^{\prime}))(1+f(p_{2}^{\prime}))$ 
which considerably increases the numerical 
efforts since in this case to evaluate the collision probability it is necessary to know the possible final 
momenta of the two colliding particles,
regardless of the fact that they actually collide.  Moreover, 
in this case the procedure of reducing the computational time by a random choice of ${\cal N}$ 
out of the possible pairs of particles 
is reasonable only choosing a large value for ${\cal N}$, otherwise the possible final states of the particles would not 
be properly mapped. In fact for $f(p)\sim f_0 \gg1$ the probability of collisions is 
enhanced with respect to the Boltzmann case by a factor $f(p)^2$. For a Bose-Einstein 
distribution $f(p\rightarrow 0) $ can be easily of the order of $10^3$ (or more)
see for example figure  \ref{f0}. This means that to keep $P^{BE}_{22}<1$ one needs 
an $\mathcal{N}$ a factor $10^6$ larger with respect to that used in the Boltzmann case to 
properly map the collisions for particle with $|\vec{p}\,|\approx 0$. This represents
the main limitation of the method proposed, which however allows to study the evolution
of overpopulated systems.   
For this reason in the following we study the evolution of the system around the onset of BEC and not at densities much above the critical one.

\subsection{Numerical checks}
In order to test the code 
we have  performed simulations in a box which allows to compare the outputs of the
 code with analytical results. We have focused mainly on two tests. One showing that 
we recover the correct equilibrium  Bose Einstein $f(p)$ and the other that the collision rate
agrees with semi-analytical estimates.
We perform therefore  simulations in  a static medium consisting of a cubic box with 
a volume $V=27$ fm$^3$
in which gluons are distributed uniformly in coordinate space, while the initial 
momentum space distribution is given by
\begin{eqnarray} 
\label{eq_glasma_f}
f( p ) = f_0 \, \theta(1 - p/Q_s )~.
\label{eq:CGC1}
\end{eqnarray}
This initial distribution is inspired by the color-glass condensate picture because it assumes that gluons are
distributed below a saturation scale $Q_s$ while modes with $p>Q_s$ are not populated;
however for the moment we consider it just a convenient initial distribution, 
with a momentum scale given by $Q_s$.
The same kind of initial distribution has been used in \cite{Blaizot:2013lga} 
for studying the evolution of a gluon gas in a static box towards the
BE condensate by mean of a Fokker-Planck approach, in the small angle approximation.
The parameters $f_0$ and $Q_s$ determine the density $n_0$ , one can easily find:
\begin{equation}
 n_0=f_0d_A\, \frac{Q_s^3}{6\pi^2}~,
\label{simple_density}
\end{equation}
and the energy density $\epsilon_0$,
\begin{equation}
 \epsilon_0  = f_0d_A\, \frac{Q_s^4}{  8\pi^2}~.
\label{simple_edensity}
\end{equation}
In the simulations we have considered a $Q_s=1$ GeV and different values of $f_0$ that will be specified in each case.
Because of the collisions the system should  evolve dynamically towards the equilibrium state,
which is characterized by a Boltzmann distribution in the case of Boltzmann statistics,
\begin{equation}
\label{diste}
f_{eq}(p) =  e^{-p/T},
\end{equation}
while for a BE gas it should evolve towards a BE equilibrium distribution, 
\begin{equation}
f^{BE}_{eq}(p)=\frac{1}{e^{(p-\mu)/T}-1}~.
\label{diste_BoseE}
\end{equation}

To compare the numerical code outputs with the analytical results we need to know
the equilibrium value of the temperature in terms of $f_0$ and $Q_s$ . In the Boltzmann case the temperature which appears in Eq.~\eqref{diste} 
can be determined using the relation $\epsilon_0=3n_0 T$.
The case of BE statistics requires more care because  also the chemical potential appears in the
equilibrium distribution.
Generally speaking temperature and chemical potential 
have to be determined solving the system
\begin{eqnarray}
\label{system_mu_t}
&& n=\frac{d_{A}T^3}{\pi^2}Li_3(e^{-\beta\lvert\mu \rvert})=n_0 \nonumber\\
&& \epsilon=\frac{3d_{A}T^4}{\pi^2}Li_4(e^{-\beta\lvert\mu \rvert})=\epsilon_0
\end{eqnarray}
where $Li_s$ corresponds to the Jonquière's polylogarithm
\begin{equation}
Li_s(z)=\sum_{n=1}^{\infty}\frac{z^n}{n^s}~.
\end{equation}
If one considers values of $f_0$ such that
the equilibrium distribution has a condensate then $\mu=0$ at equilibrium and
one recovers the well know results 
\begin{eqnarray}
\label{system_mu_0}
n&=&d_A\frac{\zeta(3)}{\pi^2}T^3 + n_c~,
\label{eq:SBn}\\
\epsilon&=&d_A\frac{\pi^2}{30}T^4~.\label{eq:SBe}
\label{density2}
\end{eqnarray}
Here $n_c$ corresponds to the fraction of particles in the condensate;
the latter however does not contribute to the energy density. Therefore
we use $\epsilon_0$ to compute the equilibrium temperature $T$. We consider a case 
slightly above the onset of condensation (see section III): $f_0=0.16$ and $Q_s=1 $ GeV. From Eqs. 
\eqref{simple_edensity} and \eqref{density2} we have $T_{eq}=\left[ 15/(4d_A)f_0 \right]^{\frac{1}{4}}\frac{Q_s}{\pi}$ ($T_{eq}=0.280$ GeV).
We use a constant total cross section 
$\sigma=1\, fm^2$ while the differential cross section is given by
\begin{equation}
 \frac{d\sigma}{dt} = \frac{9\pi \alpha_s^2}{\left(t-m_D^2\right) ^2}\left(1+\frac{m^2_D}{s}\right)~,
\label{sigma_md}
\end{equation}
where $s,t$ are the Mandelstam variables; 
such kind of cross sections are those typically used in parton cascade approaches 
\cite{Zhang:1999rs,Molnar:2001ux,Ferini:2008he,Greco:2008fs,Xu:2004mz,Xu:2008av} and by symmetry the u-channel is included.

\begin{figure}[t!]
\begin{center}
\includegraphics[scale=0.33]{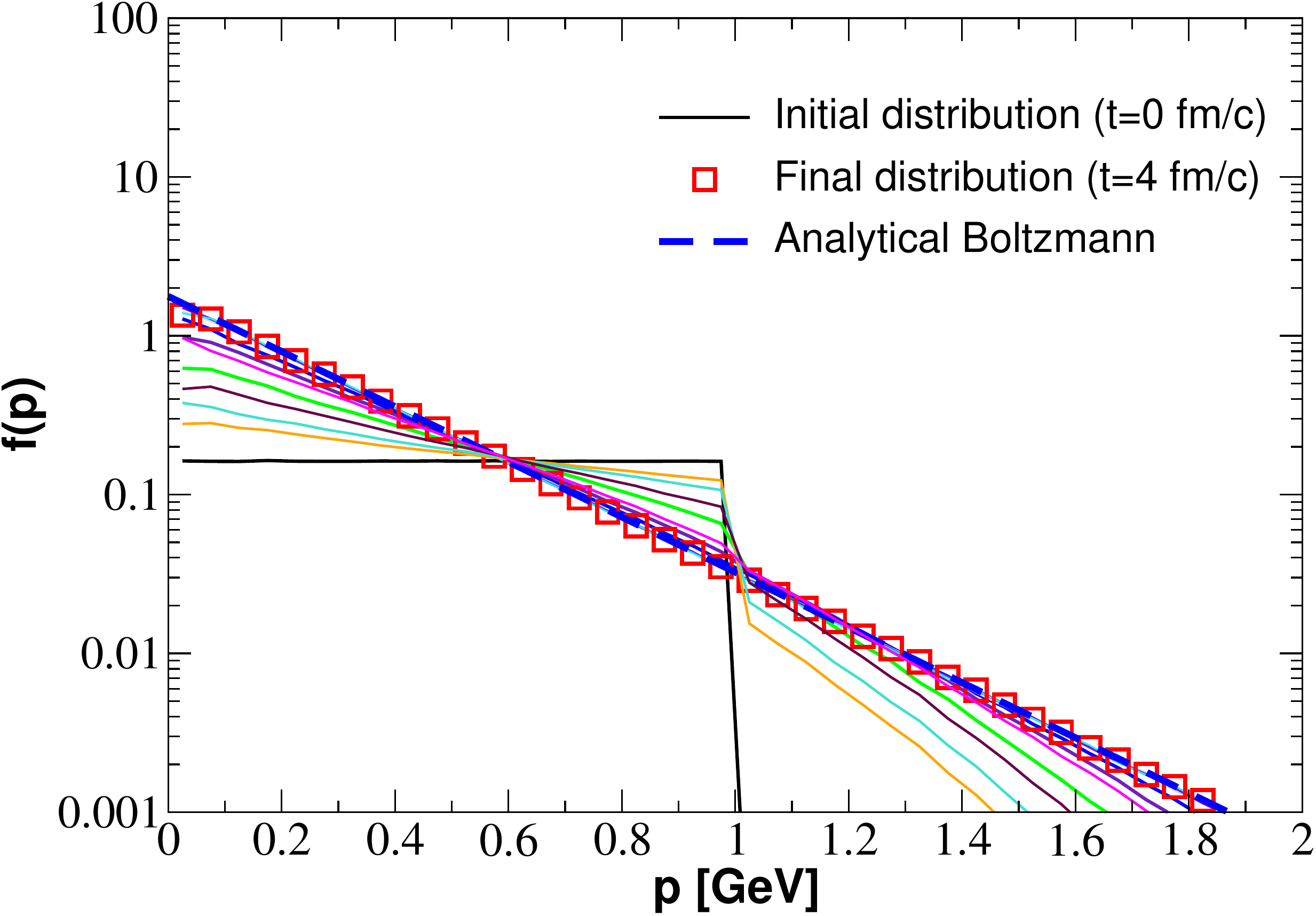}
\caption{Time evolution of the parton distribution function $f(p,t)$ without taking into account (1+f)
terms in the Boltzmann equation taken from $t=0$ every 0.5 fm/c. The initial density has been adjusted to have $f_0=0.16$. The equilibrium
distribution (open squares) is compared to the analytical expected Boltzmann equilibrium distribution (dashed curve).}
\label{fig_fp1}
\end{center}
\end{figure}

\begin{figure}[t!]
\begin{center}
\includegraphics[scale=0.3]{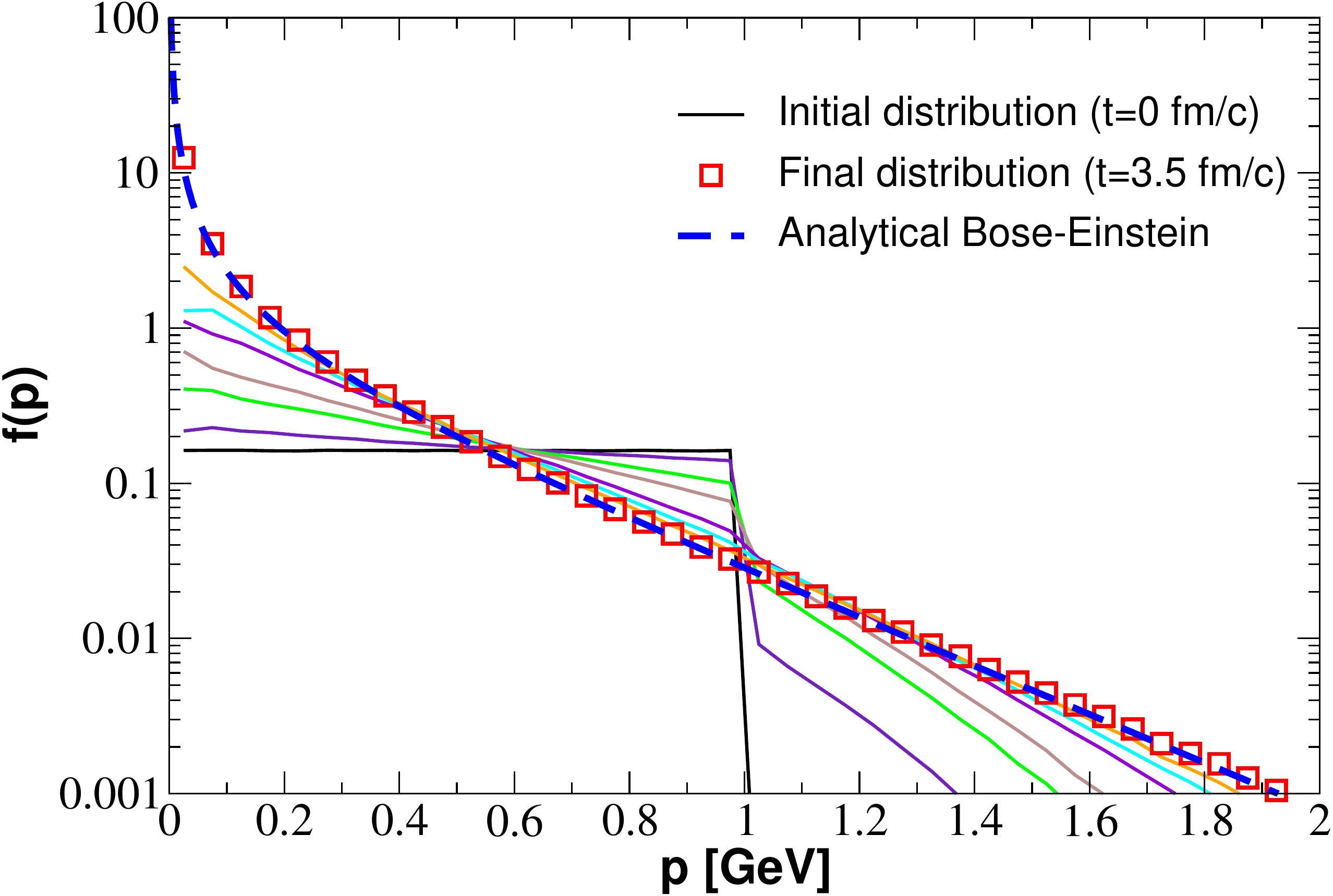}
\caption{The evolution of $f(p,t)$ for $f_0=0.16$ considering the effects of (1+f) terms in the Boltzmann equation taken from $t=0$ every 0.5 fm/c .
The distribution evolves to the correct theoretical Bose-Einstein distribution (dashed line).}

\label{fig_fp2}
\end{center}
\end{figure}

In figure \ref{fig_fp1},  we show the evolution of the momentum distribution 
obtained by the kinetic equation with the Boltzmann kernel of Eq. \eqref{CollisionInt0}, 
while in figure \ref{fig_fp2} we plot the same quantity for the case 
in which we solve the kinetic equation with a BE kernel of Eq. \eqref{one_plus_f} .
The $f(p)$ is evaluated by means of a momentum grid  $\Delta p=0.05$ GeV.
In both cases the system equilibrates (open squares)
towards the expected thermal distribution
with the proper temperature indicated by dashed line,
which is a consistency check of our simulation code.

\begin{figure}[htb!]
 \begin{center}
  \includegraphics[scale=0.3]{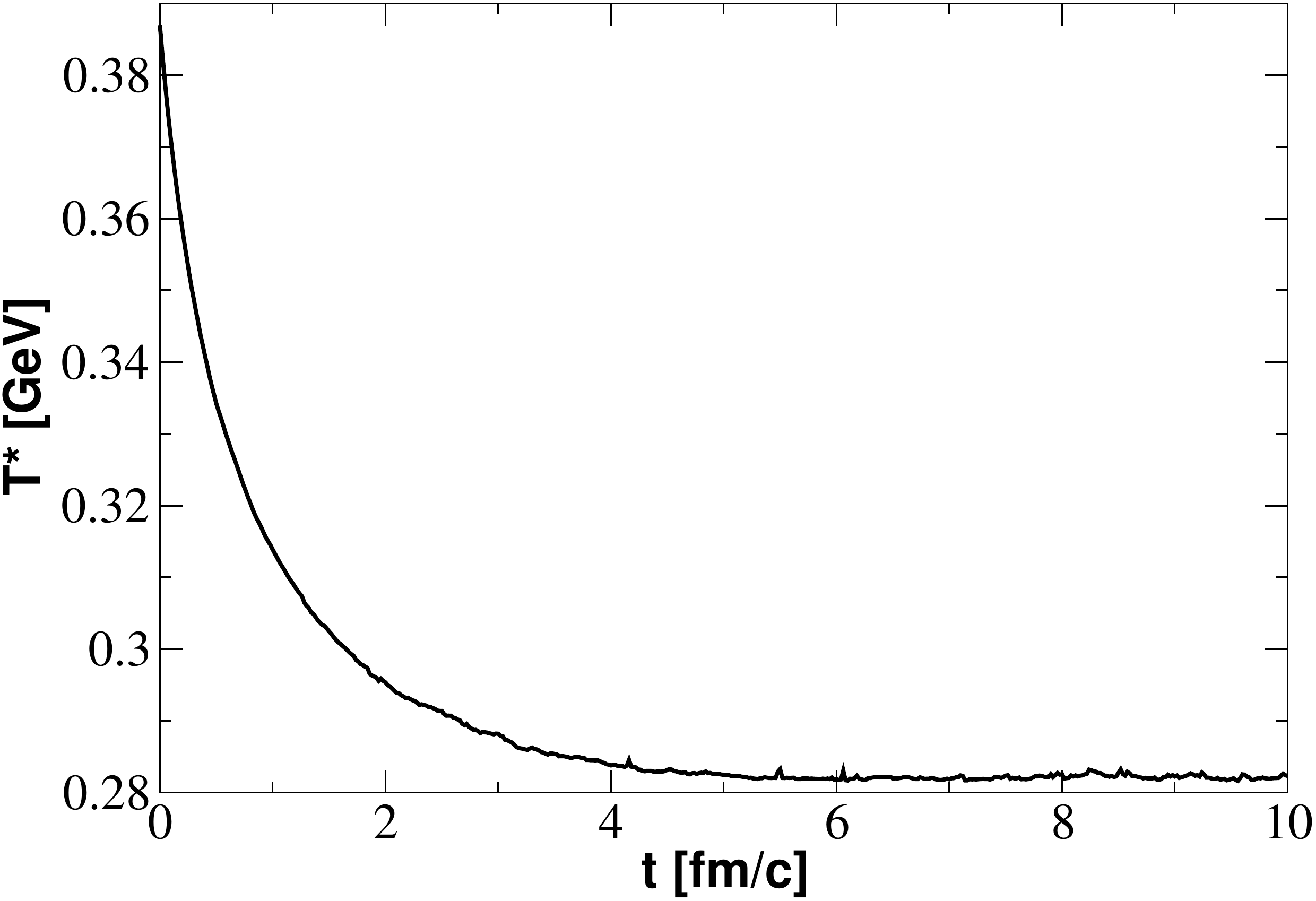}
  \caption{\label{T_gradino}Time evolution of the effective temperature $T^*$ for $f_0=0.16$, $\sigma=1\ \rm{fm^2}$ and $m_D=0.1$ GeV.}
 \end{center}
\end{figure}

Another useful numerical check is to follow the evolution of the effective temperature, $T^*$, of the system
from the initial value up to the equilibrium one.
To this end, using the notation of \cite{Blaizot:2013lga} we define
\begin{equation}
 I_a\equiv\int \frac{d^3p}{(2\pi)^3}f(\boldsymbol p)(1+f(\boldsymbol p)), \quad I_b\equiv\int\frac{d^3p}{(2\pi)^3}\frac{2f(\boldsymbol p)}{p},
\end{equation}
which when $f(p)$ is the equilibrium distribution function satisfy the relation
\begin{equation}
 I_a=T_{eq}I_b,
\end{equation}
where $T_{eq}$ is the equilibrium temperature.
Relaxing for a moment this constraint, one can define an effective temperature to be 
\begin{equation}
T^*=\frac{I_a}{I_b}~,
\label{eq:eff_tempe}
\end{equation}
at each time step of the simulation, from initial time till equilibrium,
noticing that when $f(p,t)$ reaches equilibrium it is $T^*=T_{eq}$. 
Fig. \ref{T_gradino} displays the effective temperature $T^*$ as a
function of time for the cases tested with $f_0=0.16$. 
The effective temperature in the initial stage of the evolution is larger than the equilibrium value,
then it decreases regularly until thermalization is achieved, 
and the asymptotic temperature obtained by the code coincides with the $T_{eq}$
we evaluated analytically within $0.5\%$.

\begin{figure}[htb!]
 \begin{center}
  \includegraphics[scale=0.3]{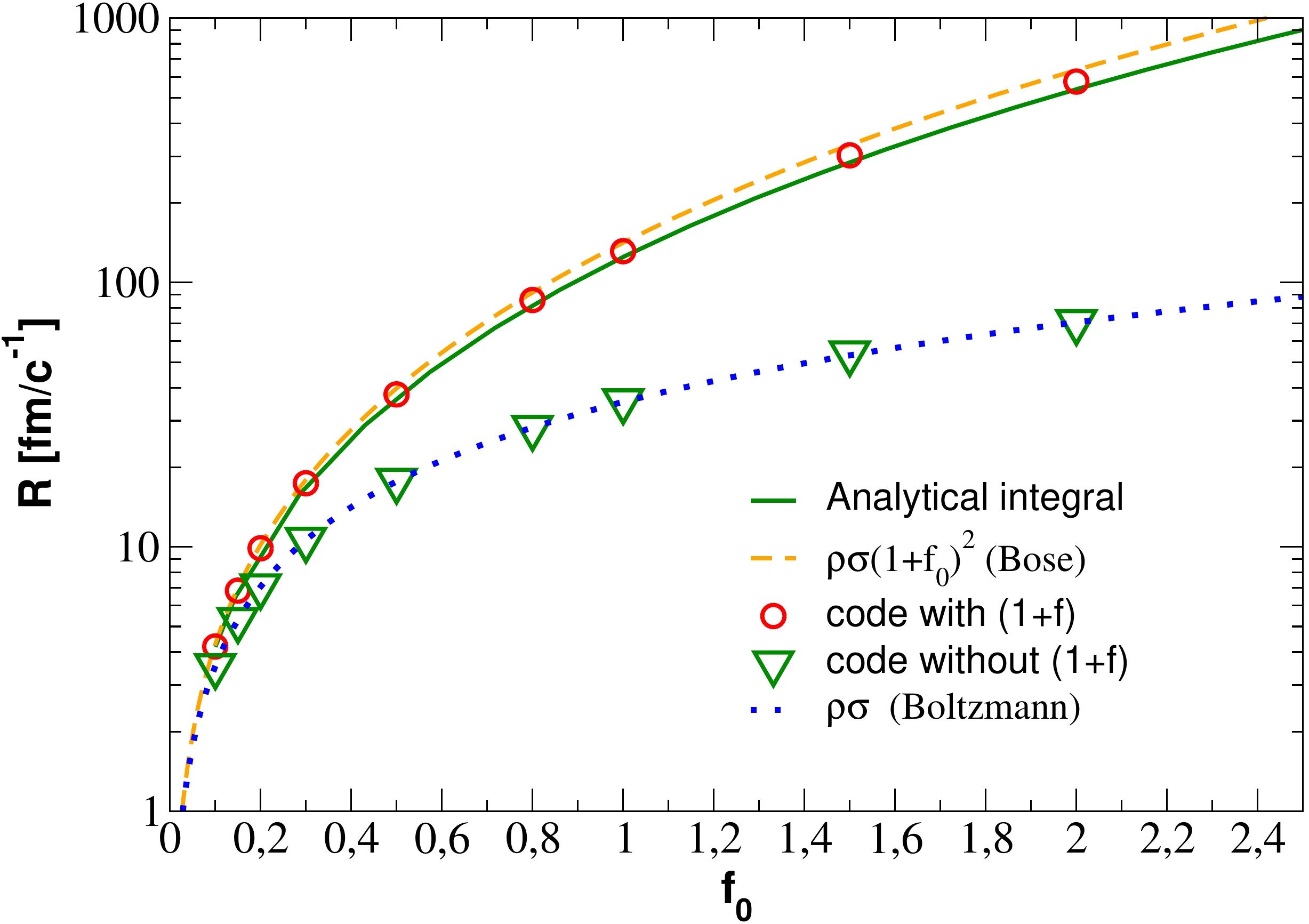}
  \caption{\label{Rate1} Collision rate per particle as a function of $f_0$.}
 \end{center}
\end{figure}

In order to make a further check of the code we have evaluated the collision rate 
$\Gamma$ 
per particle, which in the standard Boltzmann case  for a system of identical massless particles
 is simply given by $\Gamma_{\mathrm{Boltz}}=\rho^2\sigma/2$. Instead if the BE statistics is considered then the 
explicit expression for $\Gamma_{\mathrm{BE}}$ is the one derived in Appendix A and given by
\begin{eqnarray}
&&\Gamma_{BE} =\frac{1}{\nu}\int \frac{d^3p_1}{ (2\pi)^3} \int \frac{d^3p_2}{(2\pi)^3}f(p_1) f(p_2)\nonumber  \\
&& \times \int d\Omega \frac{d\sigma}{d\Omega}(1+f(p{^\prime}_1))(1+f(p{^\prime}_2))v_{rel}~.
\label{rate_cons1_text}
\end{eqnarray}
 While in the Boltzmann case the value of the 
rate for a static medium is constant during the entire evolution, since it depends only on the density and on the cross section,
in the BE case the rate 
depends on $f$ and thus it changes while $f$ evolves from
the initial non-equilibrium condition to the equilibrium one. 
We have evaluated first of all the collision rate 
at initial time, where the expression
 for the $f(p)$ is the one in equation (\ref{eq_glasma_f}), for different values of $f_0$ (densities). In particular we have
evaluated the collision rate per particle $R=2\Gamma/\rho$ and the results 
are shown
in figure \ref{Rate1}.

\begin{figure}[htb!]
 \begin{center}
  \includegraphics[scale=0.3]{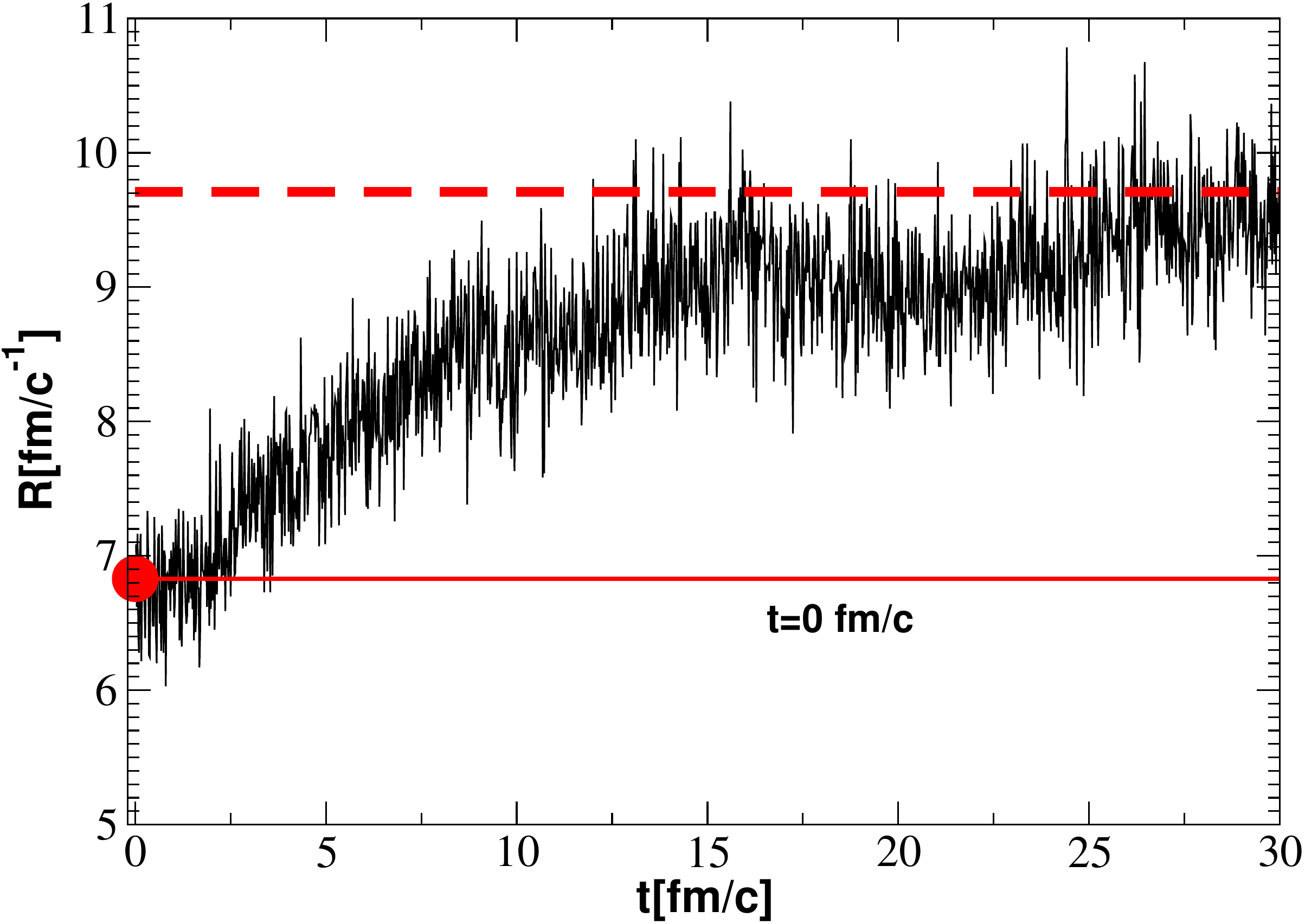}
  \caption{\label{Rate2}Rate per particle as a function of time for $f_0=0.15$. The solid line is the expected initial rate. The dashed line the one at equilibrium.}
 \end{center}
\end{figure}

It is possible for the initial time, being the $f$ a step function, to
derive  an approximate analytical expression for the collision rate that can be useful to have an idea of $\Gamma$   without
evaluating the full integral in Eq. (\ref{rate_cons1_text}).
In fact  if $f(p)=f_0$, it  can be considered constant in the whole phase space that can be explored by the system, 
 then the term $(1+f(p_{1}^\prime))(1+f(p_{2}^\prime))$ appearing
in Eq. (\ref{rate_cons1_text}) is equal to
$(1+f_0)^2$ and  Eq. (\ref{rate_cons1_text}) gives: 
\begin{equation}
\Gamma_{\mathrm{BE}} =\frac{1}{\nu}\rho_1 \rho_2 \sigma(1+f_0)^2 
\end{equation}
that for identical particle becomes 
\begin{equation}
\Gamma_{\mathrm{BE}} =\frac{1}{2}\rho^2 \sigma(1+f_0)^2=\Gamma_{\mathrm{Boltz}}(1+f_0)^2
\label{rate_approx}
\end{equation}
The approximation used to derive Eq. (\ref{rate_approx}) is that also after the collision
the $f(p)$ remains a step function, which is not exactly true because also states 
at $p>Q_s$ can be occupied after the scattering. However especially for forward peaked
$\sigma$ (small $m_D$) at $t=0^+$ it turns out to be a quite good approximation.
In figure \ref{Rate1} it is shown by the dashed line the collision
 rate for different $f_0$ evaluated using the approximate expression in Eq. (\ref{rate_approx}). 
The latter is very similar to that one evaluated solving the full integral in equation (\ref{rate_cons1_text}) indicated by the solid line 
for $m_D$ equal to $0.1$ GeV and $\sigma=1\ \mathrm{fm^2}$. The open circles in the same figure indicate the rate that we get with the code in the BE case, 
while the down triangles indicate the results we get in the Boltzmann case compared with the expected one depicted as dotted line. 
In figure \ref{Rate2} the time evolution of the rate for $f_0=0.15$ is shown. 
The circle at $t=0\ \mathrm{fm/c}$ indicates the rate evaluated using the expression (\ref{rate_cons1_text}) 
at initial time while the 
dashed line indicates the rate calculated at equilibrium through Eq. (\ref{rate_cons1_text}) 
at equilibrium.

However we warn that for $f_0>0.5\div1$ even if the rate is the expected one, the equilibrium distribution $f(p)$ for $p\rightarrow0$ is difficult to exactly map unless
a very large $\mathcal{N}$ is implemented, as discussed at the end of Sec. II.A.

\section{Initial Condition and Dynamics}
In this section we specify the initial condition we use in our simulations, in order to 
study the evolution of the system towards a thermalized state which, depending on the
initial particle and energy densities, might be a BE condensate. For what concerns coordinate space we distribute particles
 uniformly in a cubic box with a volume 
$V=27$ fm$^3$.

It is possible to estimate if the initial conditions can lead to the onset of
 a BE condensate.
As anticipated in the previous section we initialize the system by means of an out of equilibrium
distribution function which is inspired by the color-glass condensate picture, in which
gluons are assumed to populate all the momenta states below the saturation scale $Q_s$ while
states with $p>Q_s$ are empty.
The idea is as follows: given the distribution~\eqref{eq:CGC1} 
we can compute the initial particle
and energy density, which are given by Eqs.~\eqref{simple_density} and~\eqref{simple_edensity} respectively.
By means of these quantities we evaluate the dimensionless number $n_0/\epsilon_0^{3/4}$, 
introduced also in~\cite{Blaizot:2013lga} given by
\begin{equation}
\frac{n_0}{\epsilon_0^{3/4}} = f_0^{1/4}\frac{(8\pi^2)^{3/4}}{12\pi^2}~;
\label{eq:dimen_1}
\end{equation}
The same quantity is evaluated for a massless ideal boson gas at temperature $T$ and $\mu=0$
(i.e. at the onset of the BE condensation) using Eqs.~\eqref{eq:SBn} and~\eqref{eq:SBe}
with $n_c=0$:
\begin{equation}
\frac{n}{\epsilon^{3/4}} = \frac{\zeta(3)}{2\pi^2}(30\pi^2)^{3/4}~.
\label{eq:dimen_2}
\end{equation}
Comparing Eqs.~\eqref{eq:dimen_1} and~\eqref{eq:dimen_2} we find 
the value of $f_0$ which triggers the BE condensation.
Independently on the value of $Q_s$,  $f_0^{cr}=0.154$. Therefore it  is only 
$f_0$  that plays a key role in the evolution towards a Bose condensation.
In the actual simulations we slightly modify the initial distribution function
by adding an exponential decrease for $p>Q_s$ which smoothly connects the small momenta distribution
with the large momenta one. The main reason is just to have a more direct connection to \cite{Blaizot:2013lga}.
Therefore we have:
\begin{eqnarray}\label{finitCGC}
&& f(\tau_0,p) =f_0\times \nonumber\\
&&\times \, \left[ \theta(1 - p/Q_s ) + \theta(p/Q_s-1)e^{-a\, (p/Q_s-1)^2} \right]~,
\label{eq:PPPqqq}
\end{eqnarray}
which is continuous and with smooth derivative at $p=Q_s$; we choose $a=10$
and $Q_s=1$ GeV.
Initial particle and energy densities as a function of $Q_s$ are now more complicated
than equations \eqref{simple_edensity} and \eqref{simple_density}; 
nevertheless, following the same procedure pictured above, 
it is still possible to compute numerically the critical value of $f_0$ for the onset of BE condensation: 
we find $f^{cr}_{0}=0.1675$.

We will consider both $f_0$ smaller than $f_0^{cr}$, which we refer to as the underpopulated case,
and $f_0$  larger than $f_0^{cr}$ that we call the overpopulated cases. 
The total cross section corresponding to Eq. (\ref{sigma_md}) is $\sigma_{tot}=9\pi \alpha_s^2/(m_{D}^2)$.
In our calculation two quantities are left as free parameters: $m_D$ and $\alpha_s$.
In particular we consider here two values of $m_D$, namely $m_D=0.1$ GeV and $m_D=1$ GeV:
the former corresponds to a forward peaked cross section, that justifies a small angle
approximation where the kinetic equation should reduce to a Fokker-Planck evolution~\cite{Blaizot:2013lga}.
The larger value of $m_D$, which amounts to a magnitude relevant for relativistic heavy ion collisions, 
corresponds to a more isotropic cross section $(m_D/T\cong2)$: in this case the small 
angle approximation could be 
no longer an accurate approximation of the kinetic equation \cite{Das:2013kea}.
Once we fix $m_D$ in the calculation, the remaining parameter is $\alpha_s$ that fixes the value of
the total cross section.
We are interested to employ cross section that can be of the order of those supplying  
an $\eta/s \approx 0.1$. We use the following approximate relation to choose the value of $\sigma_{tot}$ \cite{Plumari:2012xz,Plumari:2012ep}:
 $\sigma_{tot}=\frac{3}{10}T_{eq}/(n\eta/s)$.
The value of $\sigma_{tot}$ depends through $n$ on
the value of $f_0$. For $f_0=0.1;0.2;0.3;0.5$ the cross sections are respectively $\sigma_{tot}=0.7;0.38;0.28;0.19$ $\mathrm{fm^2}$.
This formula has to be considered as a rough approximation, in fact for 
most of the evolution one does not have really a temperature $T_{\mathrm{eq}}$ and we are discarding 
the difference between the transport and the total cross section. Our aim is mainly 
to study the impact of $m_D$ on the dynamics toward the BE condensation at fixed cross section to understand the effect of the large angle scattering. 

Another approach is using the one-loop $\beta$-function to compute $\alpha_s$ at a given temperature, then using the  
relation $m_{D}^{2}=4\pi\alpha_sT^2$ to evaluate the Debye mass. This results in a large increase of the cross
section with increasing $m_D$. We will discuss more about this in section V.

\section{Thermalization in the under-populated case}
We discuss now the thermalization dynamic from the CGC inspired $f(p)$ to the BE 
distribution function. 
We have discussed that with the initial condition
specified by Eq.~\eqref{eq:PPPqqq} it is expected that the systems evolves towards a BE condensate,
depending on the value of $f_0$: for $f_0<f_0^{cr}$ the system equilibrates towards a BE distribution
with a finite chemical potential;
for $f_0>f_0^{cr}$ a fraction of particles forms a BE condensate
and the equilibrium distribution is characterized by a vanishing chemical potential.
In this section we focus
on the case $f_0<f_0^{cr}$, i.e. the case in which the system does not reach the condensate phase.

\begin{figure}[t!]
 \begin{center}
  \includegraphics[scale=0.3]{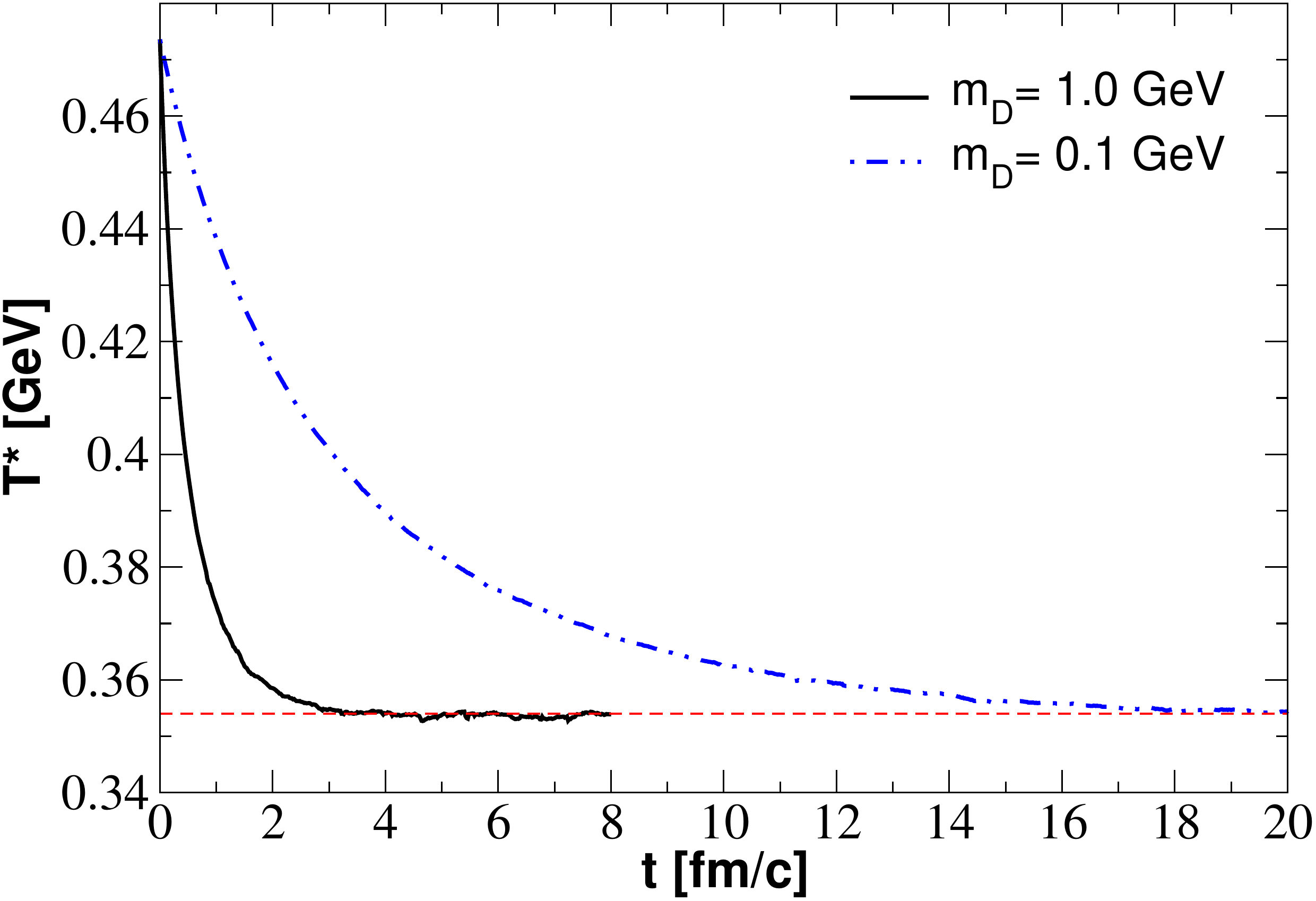}\\
   \includegraphics[scale=0.3]{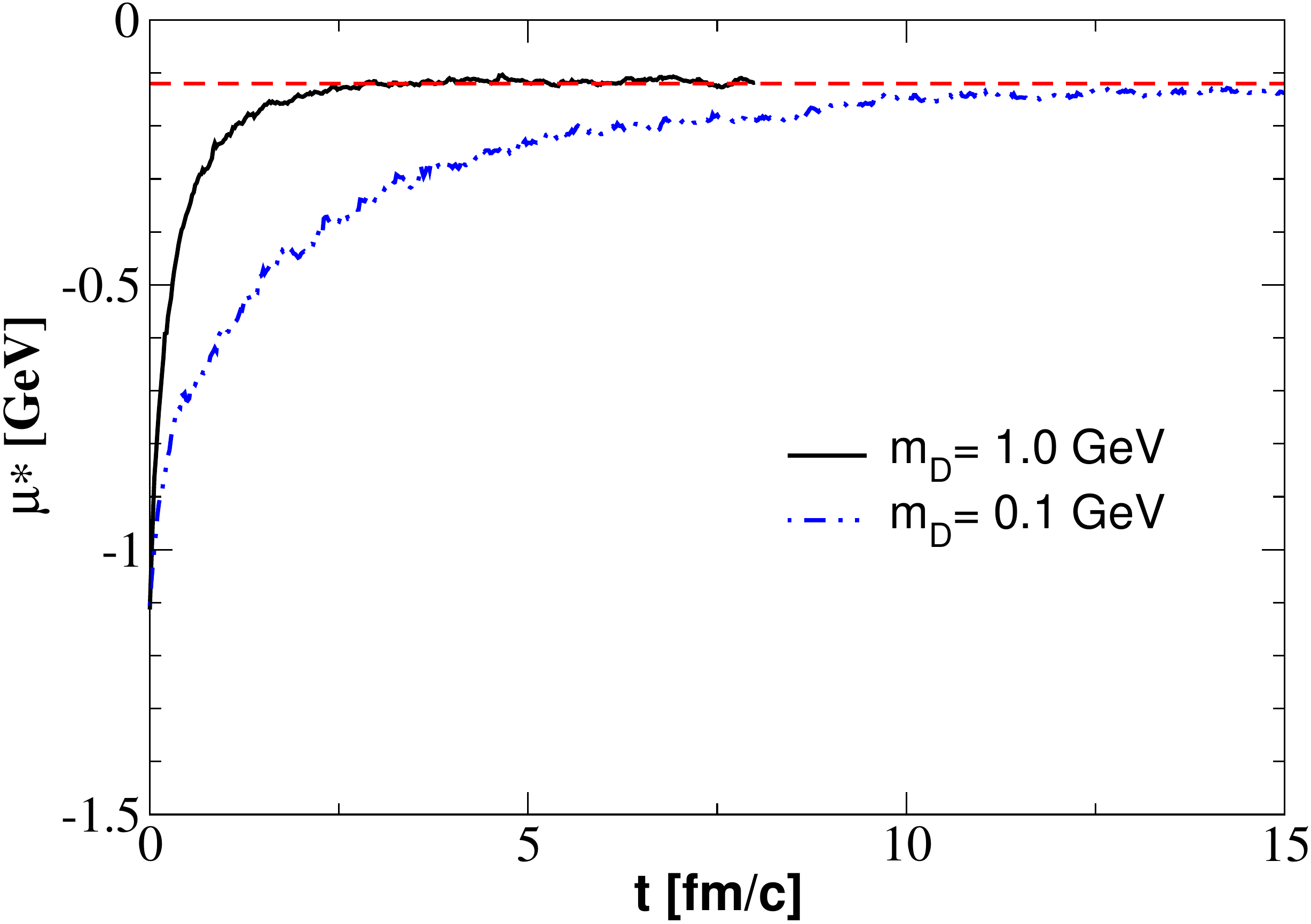}
  \caption{\label{T_under}{\em Upper panel.} Time evolution of the effective temperature $T^*$ in the under-populated case ($f_0=0.1$) for
$m_D=0.1$ GeV (top curve) and $m_D=1$ GeV (bottom curve). The predicted equilibrium value $T^*=0.354$ GeV is represented by the horizontal dashed line.
{\em Lower panel.} Time evolution of the effective chemical potential for $m_D=0.1$ GeV (bottom curve) and $m_D=1$ GeV (top curve).}
\label{mu-evol}
 \end{center}
\end{figure}

In the upper panel of figure \ref{T_under}, we plot the effective temperature $T^*$,
defined by Eq.~\eqref{eq:eff_tempe}, as a function of time for $f_0=0.1$; 
the two curves correspond to the same value of the total cross sections,
while the numerical value of the screening mass is different: dashed line corresponds to $m_D=0.1$
 GeV,
while solid line to $m_D = 1$ GeV. The initial $T^*$ is of course independent on $m_D$.
In both cases the effective temperature at initial time is larger than the equilibrium
 value and decreases smoothly until thermalization is complete.
It is clear however that time needed to achieve thermalization is considerably affected 
by $m_D$, that is by the anisotropy of the cross section.
In fact, whereas for the case of the forward peaked cross section
thermalization occurs in about $\tau_{therm}\approx 15$ fm/c, 
the equilibration time for $m_D=1$ GeV is much smaller, $\tau_{therm}\approx 3$ fm/c.

\begin{figure}[t!]
 \begin{center}
  \includegraphics[scale=0.3]{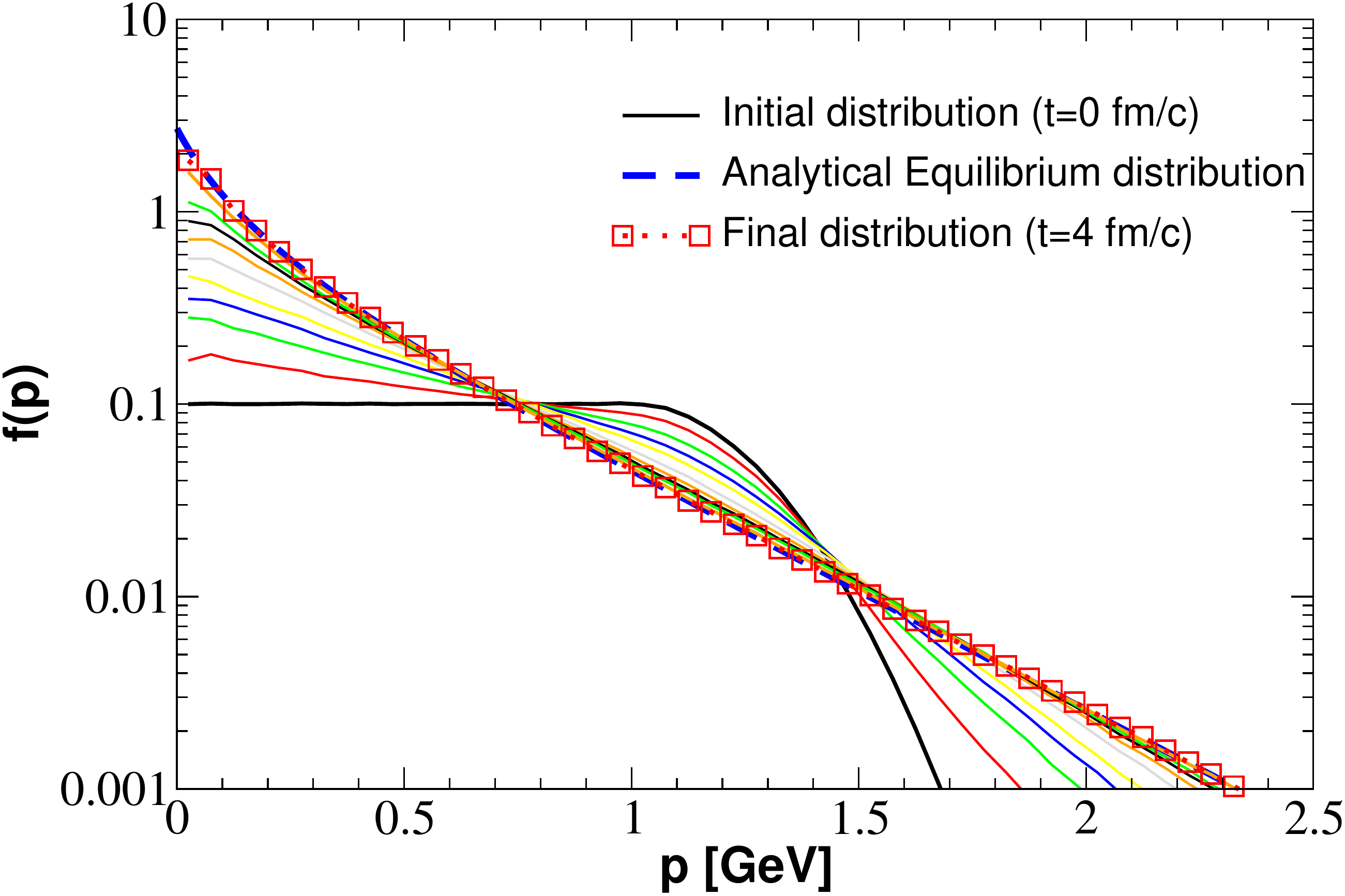}
  \caption{\label{f01}Time evolution of the distribution function $f(p,t)$ in the under-populated case ($f_0=0.1$) for
$m_D=1$ GeV. The equilibrium distribution obtained by the simulation (squares) is fitted by a Bose-Einstein distribution (dashed) with $T=0.354$ GeV and $\mu^*=-0.11$ GeV.}
 \end{center}
\end{figure}

At each time step in the simulation an effective chemical potential, $\mu^*$, can be defined
by the distribution function at $\bm p=0$,
\begin{equation}
 f(\bm p=0)=\frac{1}{e^{-\mu^*/T^*}-1}~.
\end{equation}
At equilibrium $\mu^*$ coincides with the chemical potential of the system.
Extending this approximation 
in the region $p\approx0$
we compute $\mu^*$ by 
averaging the distribution at small momenta
\begin{equation}
 \mu^*=\frac{\sum\limits_{n=0}^{k}\left[p_n-T^*\mathrm{ln}\left(1+\frac{1}{f(p_n)}\right)\right]}{k}~,
\end{equation}
where $p_n$ is the discretized value of momentum with index $n$, and $k$ is the number of points near $p\simeq0$ considered to evaluate the average.
In what follows, we used a momentum grid size of $\Delta p=0.05$ GeV and a $k=3$.
A plot of the effective chemical potential is shown in the lower panel of Fig. \ref{T_under}, 
where it can be seen that $\mu^*$ reaches smoothly its equilibrium value $|\mu^*|\simeq0.11$ GeV.
We stress the fact that at equilibrium, 
the effective temperature and chemical potential computed by the code are 
the same ones as estimated analytically and are indicated by dashed lines in Fig.
\ref{mu-evol}. The distribution function is very well fitted by a BE distribution 
with the same values of $T^*$ and $\mu^*$, as can be inferred from Fig.\ref{f01}.
A precise quantitative comparison with \cite{Blaizot:2013lga} is not direct because
there the screening mass is not specified, however the results are shown in term of
$\tau=2\pi^2 \alpha_{s}^{2}\xi f_0(1+f_0)t$, with $\xi=\frac{18}{\pi} \mathcal{L}\cong \frac{18}{\pi} \mathrm{ln}(T/m_D)\cong1$. For $f_0=0.1$ in \cite{Blaizot:2013lga} $\tau_{\mathrm{eq}}\approx 20$
which should correspond to $t\sim 13$ fm/c, in quite good agreement with our results for the forward peaked case ($m_D=0.1$ GeV).

\begin{figure}[t!]
\begin{center}
\includegraphics[scale=0.3]{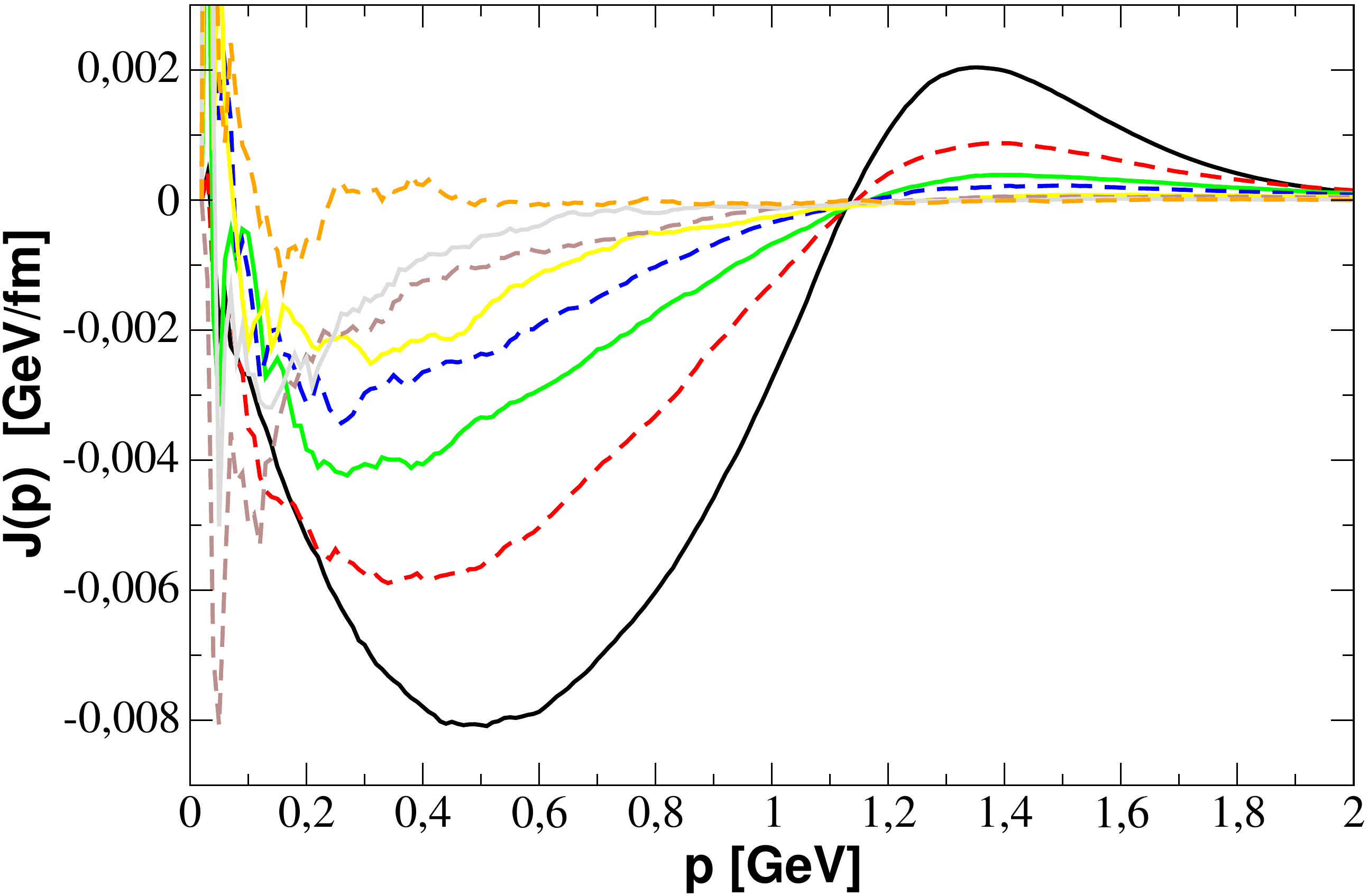}\\
\includegraphics[scale=0.3]{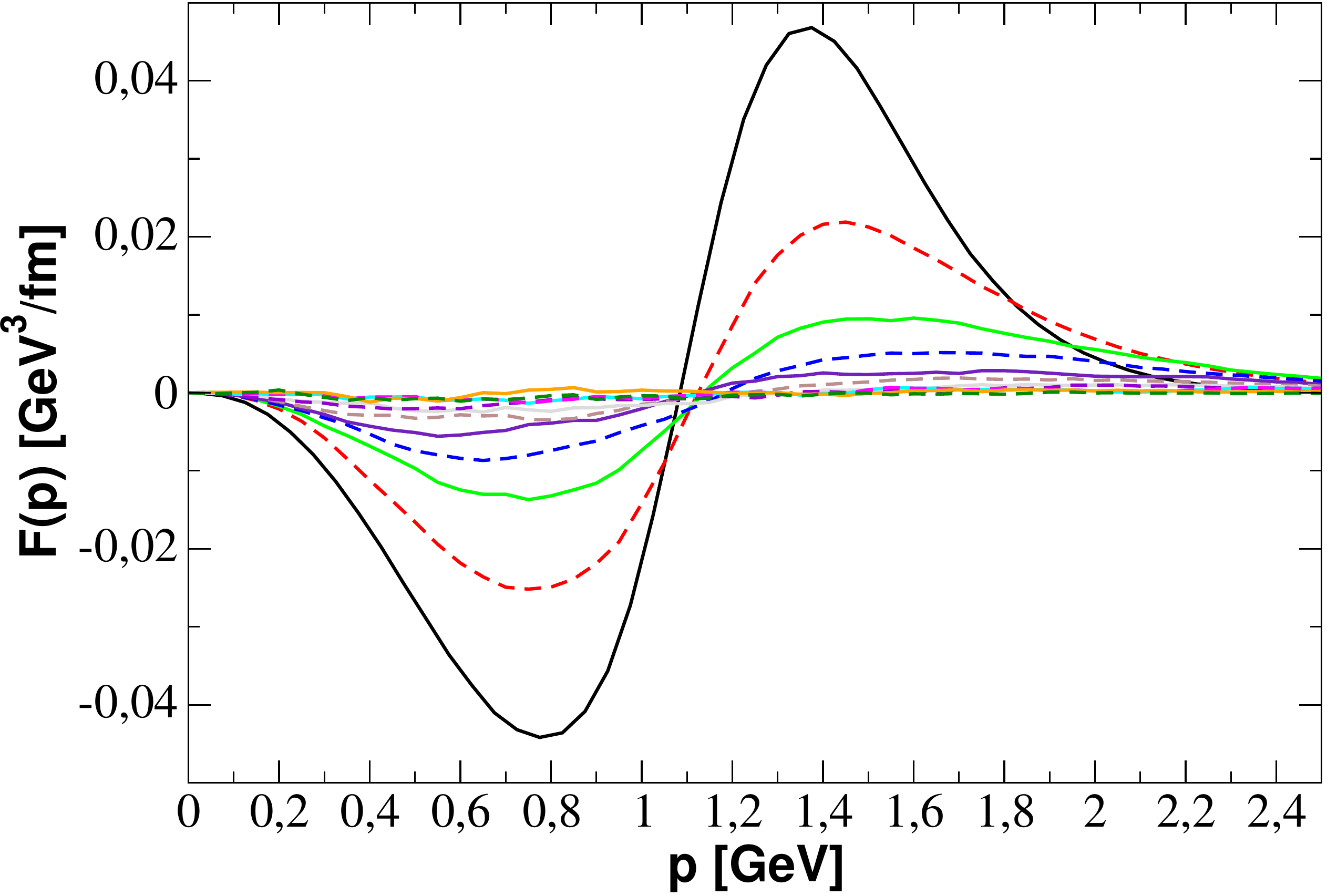}
\caption{\label{current}({\em Upper panel}). Current $\mathcal{J}(p,t)$ evolution for $f_0=0.1$ from initial time (bottom curve at low momenta) till the reaching of equilibrium. Case $m_D=1$ GeV.
Different curves are taken every $0.5$ fm/c.
({\em Lower panel}). Momentum space flux ${\cal F}(p,t)$. }
\end{center}
\end{figure}

Following \cite{Blaizot:2013lga} we introduce two further quantities, namely the flux $\mathcal{F}(p,t)$ and the current
${\cal J}(p,t)$.
We evaluate the flux at a given momentum $p$ and fixed time as defined in the follow 
\footnote{We have supposed that $\mathcal{F}(0)=0$ as it has been done in \cite{Blaizot:2013lga}.} 
\begin{equation}
\mathcal{F}(p,t)=-(2\pi)^3\frac{\Delta n_{p}}{\Delta t}~,
\end{equation}
where $n_{p}$ corresponds to the spatial density of particles in a sphere of radius $p$.
The sign convention implies that the flux is positive when there is a net flux of particles going out from
the momentum space sphere.
The current is related to the flux by the relation
\begin{equation}
 \mathcal{J}(p,t)=\frac{\mathcal{F}(p,t)}{4\pi p^2}~.
\end{equation}
The time evolution of these two quantities is summarized in Fig. \ref{current} .

In the upper panel of Fig. \ref{current} we plot our result for the current as a function of momentum magnitude
for several time steps, from initial stage up to equilibration time.
The results shown correspond to the case $m_D=1$ GeV.
The shape of the current reflects the flow of particles in momentum space during the evolution. 
Particularly noticeable is the initial growth of low momenta at the expenses
of the region near $Q_s$, behavior that confirms the tendency already seen in the evolution of the distribution function. 
As the system approaches equilibrium the distribution function stabilizes on the BE one, 
and this fact is also visible in the figure of the current, as 
it decreases in absolute value, being almost 
vanishing after a time $t\approx4$ fm/c.
We briefly comment that the fluctuations in the regime of very small momenta, namely $p\leq0.1$ GeV,
are related to those of $f(p)$ and have not to be considered 
as physical effects:
these are mainly due to numerical fluctuations when $\mathcal{F}(p)\approx0$
(lowering further such numerical fluctuations would considerably increase the computational cost).
In the lower panel of Fig. \ref{current} we plot the flux ${\cal F}(p,t)$, which shows a behavior 
very similar to that of the current. Again, the flux nearly vanishes when
the effective temperature reaches its equilibrium value (dashed curves in Fig \ref{current}).
We have checked that the picture summarized in Fig. \ref{current} does not change
qualitatively by increasing $f_0$ providing $f_0 < f^{cr}_{0}$ .
Again, we find a quite similar behavior for both $\mathcal{F}(p)$ and $\mathcal{J}(p)$ to \cite{Blaizot:2013lga} with
a very similar magnitude of the peaks and in particular a maximum absolute value of $\mathcal{J}(p)$ that is initially about 4 times larger in the 
infrared region ($p<Q_s$).

\section {Over-populated case: the onset of BEC}

As discussed in the previous section, 
when $f_0>f_0^{cr}$ the system evolves towards an equilibrium state in which a BEC is present. 
This equilibrium state is called over-populated because a BE distribution at $\mu=0$ and with equilibrium
temperature cannot accomodate all the particles and a finite fraction of them is stored in the $\bm p=0$ state
forming a condensate.
Without imposing specific boundary conditions such as a non-vanishing flux of particles at $p=0$ \cite{Blaizot:2014jna}, 
as well as introducing a coupling between the particles in the bulk and the condensate, 
one cannot monitor the system all the way up to equilibrium \cite{Semikoz:1995rd}. 
However we can still use our formalism to study the evolution of the system from the initial condition
till the onset of condensation, 
which naturally appears in our approach by the fact that $\mu^*\rightarrow 0$ in a finite
time $t_{BEC}$.

\begin{figure}[t!]
 \begin{center}
  \includegraphics[scale=0.3]{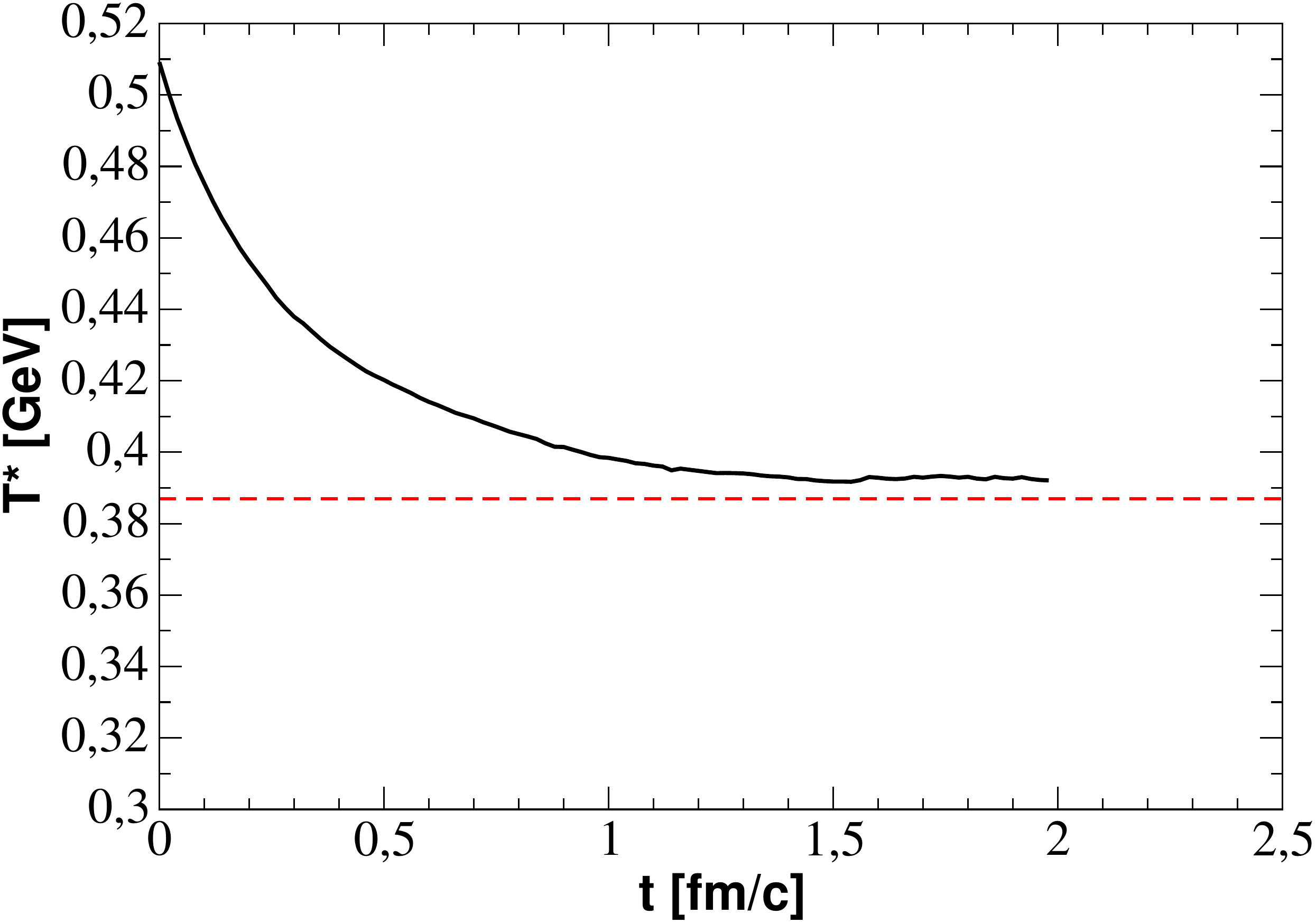}
  \caption{\label{Temp_over}The effective temperature $T^*$ as a function of time in the over-populated case ($f_0=0.2$). The dashed line represents the theoretical predicted value of 
   $T^*=0.386$.}
 \end{center}
 \end{figure}
 
We begin our analysis from the effective temperature $T^*$, displayed in Fig. \ref{Temp_over}. As was noted for the under-populated case, the effective temperature
lowers regularly in time. The peculiar characteristic of this case however lies in the fact that $T^*$ does not reach its final equilibrium value because condensation sets in before thermalization. 

\begin{figure}[t!]
\begin{center}
  \includegraphics[scale=0.3]{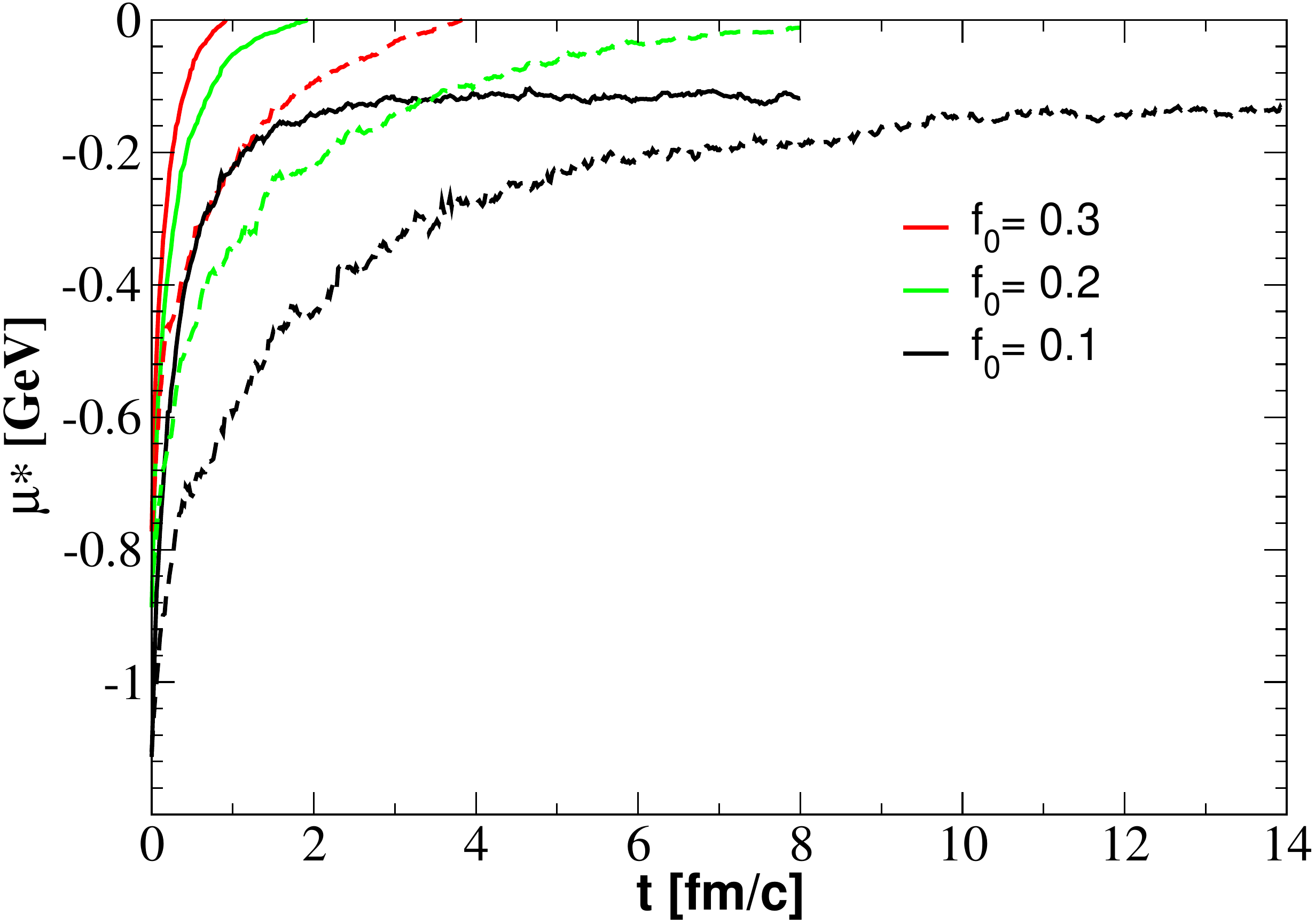}
  \caption{\label{mu}The chemical potential $\mu^*(t)$ for various initial densities (from bottom to top $f_0=0.1$,0.2,0.3). For each density two curves are showed, one for the $m_D=1$ GeV case (solid),
  the other for $m_D=0.1$ GeV (dashed).}
 \end{center}
\end{figure}

The most important signature of the transition to condensate phase is, as anticipated, the vanishing of 
the effective chemical potential $\mu^*$.
In Fig. \ref{mu}, we plot $\mu^*$ for several values of $f_0$;
solid lines are obtained for $m_D=1$ GeV, whereas dashed ones for $m_D=0.1$ GeV.
Even if we want to focus here on the over-populated case, we show also results for the
under-populated case to make a clearer comparison with the former one and
enlighten the differences among the two regimes.

Many things in this picture are noteworthy. 
Firstly, $\mu^*$ vanishes in a finite time range for all cases above the critical density, while it remains
negative for $f_0<f_0^{cr}$, and this is true independently on the choice of the angular part of the cross section
(that is on the value of $m_D$). We have also checked that $\mu^* \rightarrow 0$ exactly at $f_0^{cr}=0.1675$ with a precision of $0.5\%$
Moreover, the time $t_{BEC}$ at which $\mu^*$ vanishes depends on the density and, namely, is larger when the density is lower.
For a fixed density and $f_0>f_0^{cr}$, cases with $m_D=0.1$ GeV reaches the condensate phase about a factor 4 more slowly than $m_D=1$ GeV. 
Roughly the same factor is observed in the case $f_0<f_0^{cr}$ where the two plots reach the same equilibrium value of $|\mu^*|\simeq0.11$.




For times close to the $t_{BEC}$ one can set 
\begin{equation}
 |\mu^*|=C(t_c-t)^\eta,
\end{equation}
where $\eta$ plays the role of a critical exponent.
Fits of the numerical results with this function, keeping $\eta$ as a free parameter, as well as the values of the slope $C$ and critical time
$t_{BEC}$ as functions of $f_0$ have indicated a value of $\eta=1.3\pm0.1$ for $m_D=0.1$ GeV and $\eta=1.6\pm0.1$ for $m_D=1$ GeV.
Therefore with respect to \cite{Blaizot:2013lga} where $\eta=1$ we find a large value even for $m_D=0.1$ GeV where the soft scattering approximation should be safely applicable. 
It can be noted that, as already seen in Fig. \ref{mu}, the critical time decreases as $f_0$ increases, tending to diverge as $f_0\rightarrow f_0^{cr}$.


\begin{figure}[t!]
\begin{center}
\includegraphics[scale=0.3]{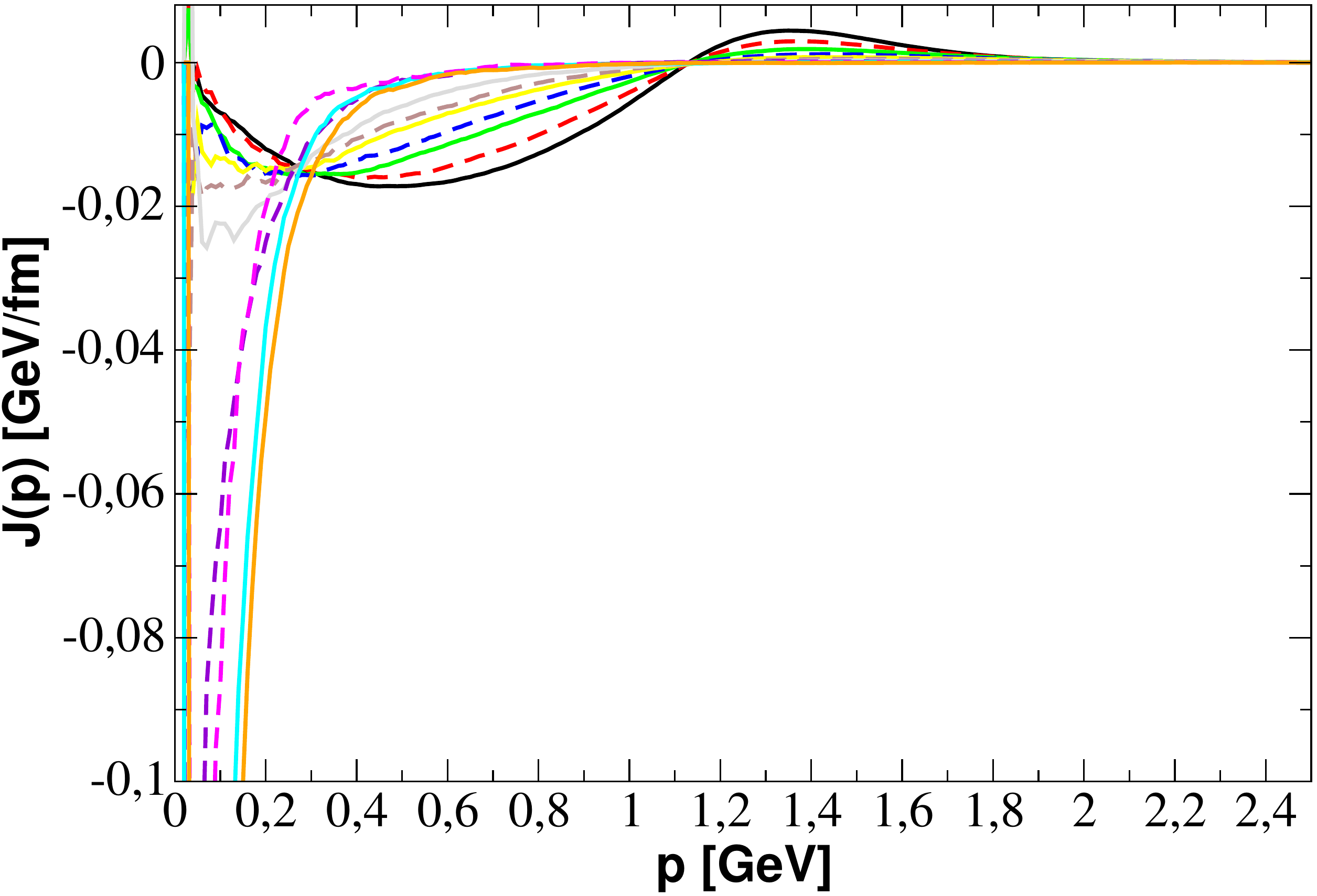}
\caption{Time evolution of the current $\mathcal{J}(p)$ in the over-populated case ($f_0=0.3$). The curves represent the currents every 0.15 fm/c.} 
\end{center}
\end{figure}
\begin{figure}[t!]
\begin{center}
\includegraphics[scale=0.3]{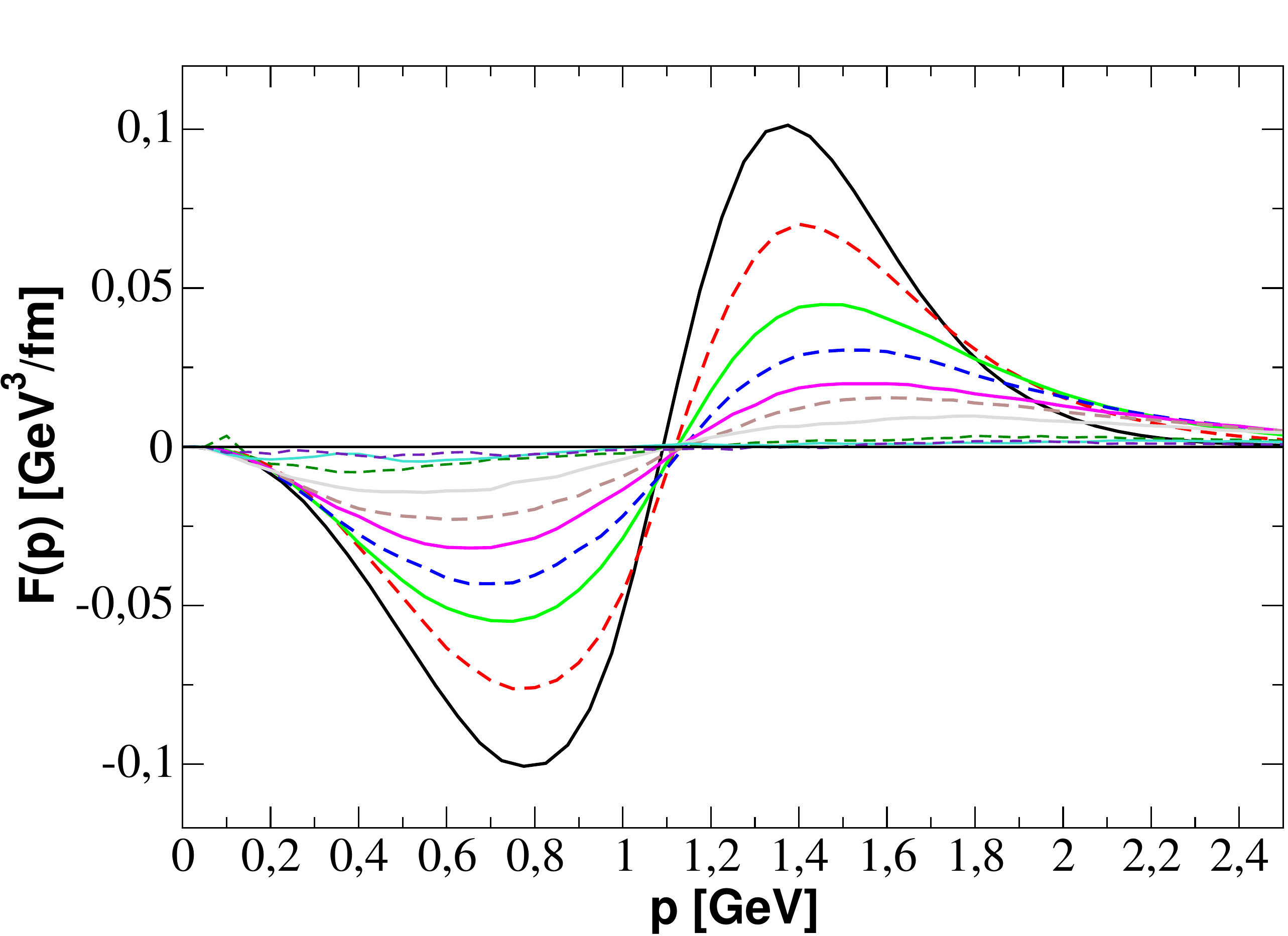}\\
\caption{The flux $\mathcal{F}(p)$ ($f_0=0.3$) for various times till the reaching of thermal equilibrium. The curves represent the flux every 0.15 fm/c.\label{Fig:flux_bec}}
\end{center}
\end{figure}

The onset of condensation is visible also in the modified behavior of the current with respect to the case treated in the previous section.
The important dynamics here is in the low $p$ region, where the current shows a strong increase in absolute value as time approaches $t_{BEC}$.
While for $f_0<f_0^{cr}$ the peak in $\mathcal{J}(p)$ appears at $p \sim0.8$ GeV, at $f_0>f_0^{cr}$ it is shifted on the $p\rightarrow0$ region.
This fact is the proof that, when the condensate phase is reached, there is a net number of particles with low momentum going from the gluon gas to the condensate itself.
This behavior is not easily recognized in the plot of the flux, Fig.~\ref{Fig:flux_bec}, 
because the increase of the current is absorbed by the factor $p^2\rightarrow0$.
Qualitatively we confirm the behavior discussed in \cite{Blaizot:2013lga}.

\begin{figure}[t!]
\begin{center}
\includegraphics[scale=0.3]{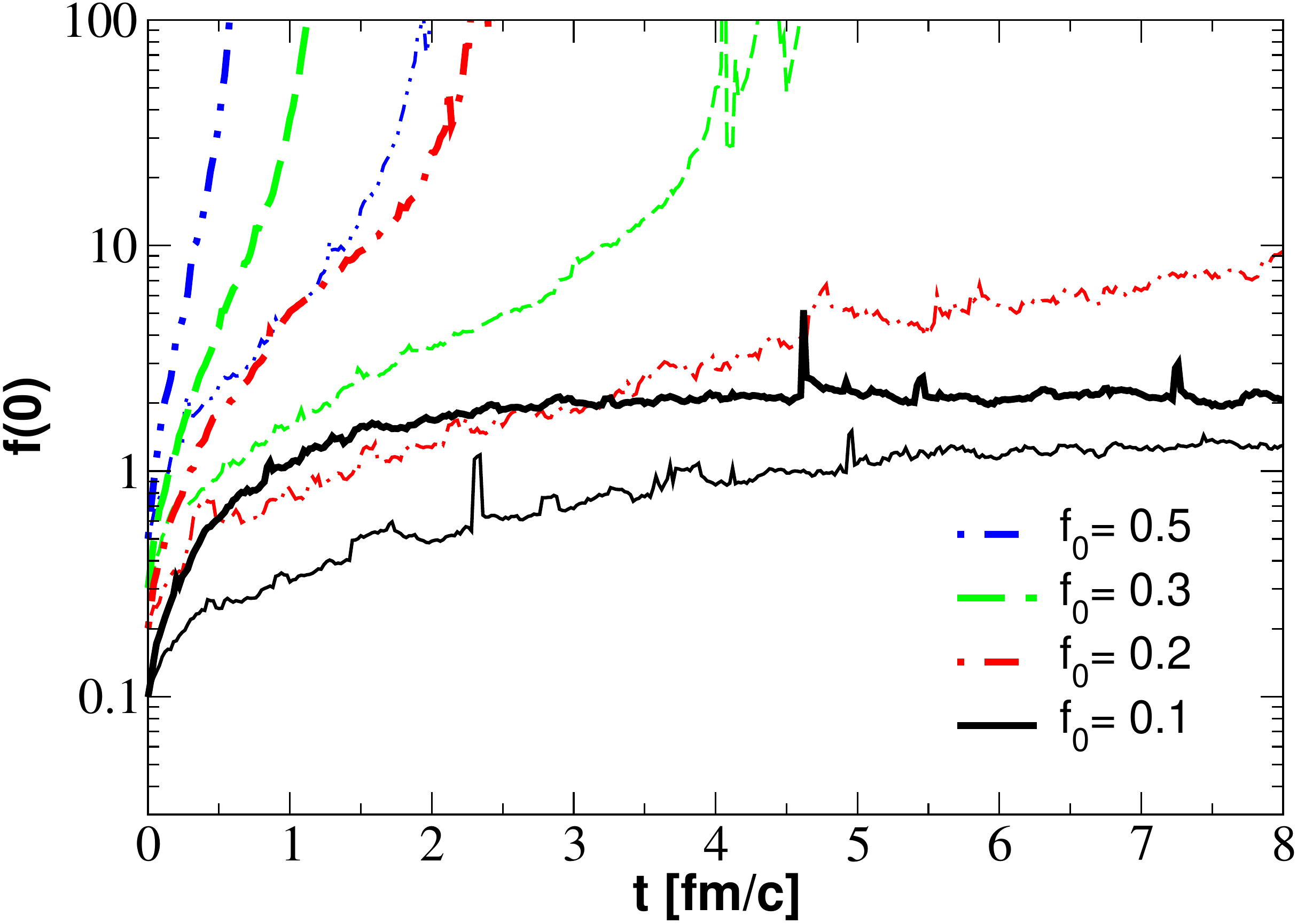}\\
\caption{$f(p\simeq0,t)$ for various initial densities and for two different gluon screening masses $m_D=0.1$ GeV (thin lines), $m_D=$1 GeV (thick lines).
From bottom to top the $f_0$ are: 0.1, 0.2, 0.3, 0.5.}
\label{f0}
\end{center}
\end{figure}

Finally, in Fig. \ref{f0} a plot of $f(p\simeq0,t)$ is presented.
The difference between over-populated and under-populated cases 
is clearly seen. As a matter of fact, while the $f_0=0.1$ curve rises slowly and saturates when thermal equilibrium is reached,
in the case $f_0>f_0^{cr}$ there is a huge increase of $f(p\simeq0,t)$ without any saturation: $f(p,t)$ develops in fact a singularity at $p=0$. 
Again we see that the $m_D=0.1$ GeV case (forward peaked) is quite 
slower than the $m_D=1$ GeV case. The latter however should be more close to the $m_D$
in a QGP medium at $T\sim 0.4-0.5$ GeV as those explored at LHC energy.

\subsection{Perturbative case}
In the final part of this article we report also our results obtained assuming
a perturbative QCD dynamics still governed by elastic two body collisions with
cross section given by~\eqref{sigma_md}, but instead of fixing by hand $m_D$ and $\alpha_s$
we have considered a running QCD coupling  
\begin{eqnarray}
\alpha_s(Q^2)=\frac{4\pi}{\beta_0\, \text{ln}(-Q^2/\Lambda_{QCD})}~,
\label{eq-alpha}
\end{eqnarray}
where $\beta_0=11-\dfrac{2}{3}N_f $ and the thermal scale $Q^2=(2 \pi\,T)^2$, with Debye screening mass 
$m_D=g(T)T$ where $g(T)=\sqrt{4\pi\alpha_s(T)}$. Since we consider a system made of only gluons we put $N_f=0$
in the above equation.
As a reference temperature to carry out calculations for $\alpha_s(T)$ and
$m_D(T)$, the effective temperature $T^*$ has been used.


\begin{figure}[t!]
\begin{center}
  \includegraphics[scale=0.3]{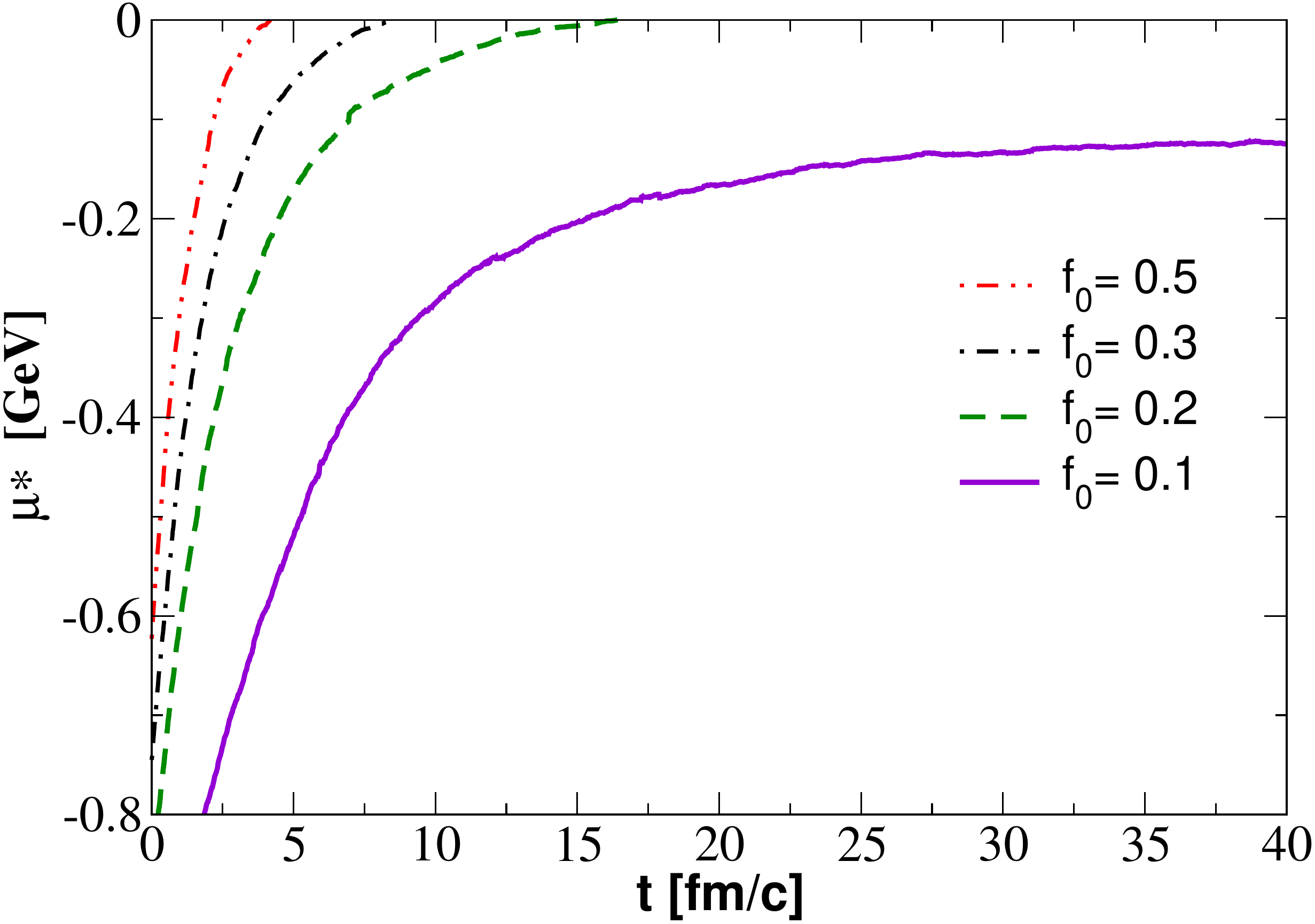}
  \caption{\label{mu_p}The chemical potential $\mu^*$ for various initial densities in the perturbative case.}
 \end{center}
\end{figure}

In Fig. \ref{mu_p} we plot the results for the effective chemical potential in this calculation.
In analogy to what we observed in the previous cases, for $f_0>f_0^{cr}$ the effective chemical potential
reaches zero in a finite amount of time, signaling the onset of the BE condensation,
while for $f_0<f_0^{cr}$ it evolves to a nonvanishing value  which is independent on the cross section,
see also Fig. 10.
\begin{figure}[t!]
\begin{center}
\includegraphics[scale=0.3]{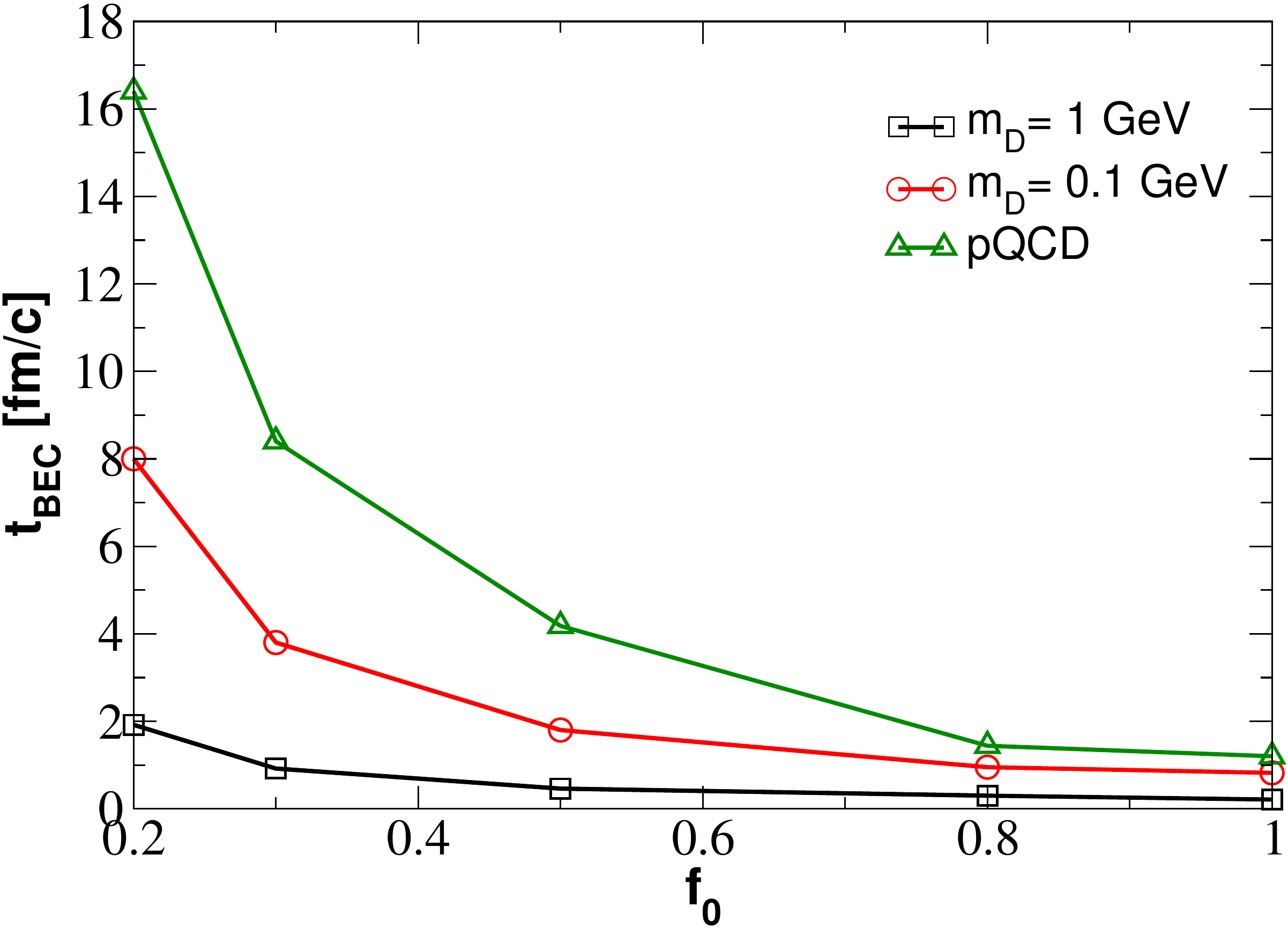}\\
\caption{Variation of the critical time $t_{BEC}$ at which the system undergoes the transition to the condensate phase as a function of initial density.
Both $m_D=1$, 0,1 GeV cases are showed.}
\label{tc_f0}
\end{center}
\end{figure}
In Fig. \ref{tc_f0} we summarize the $t_{BEC}$ as a function of $f_0$ for the different
cases considered. It is important to have an approximate time scale under the 
typical condition at uRHIC's. This can be done noticing that $f_0=0.3$ corresponds
to $n\simeq23\ \mathrm{fm}^{-3}$ which is the density at the center of the fireball at RHIC energies
at $\tau \simeq 0.5\ \mathrm{fm/c}$ while $f_0=0.8$ means a $n\simeq 60\ \mathrm{fm}^{-3}$ that is roughly the maximum density reached at LHC energy. 
At $f_0=0.3$ with an $m_D=1$GeV we see that
$t_{BEC}\simeq 1\ \mathrm{fm/c}$. For RHIC conditions this means that a dynamical BEC 
can be hardly reachable, considering also the strong longitudinal expansion. 
However for LHC condition $t_{BEC}\lesssim 0.2\ \mathrm{fm/c}$ at $t\approx 0.5\ \mathrm{fm/c}$ which means 
that there could be the condition to observe at least a transient BEC. It is also important
to notice that for the pQCD case the $t_{BEC}$ is generally quite large respect
to the expansion rate of uRHIC's. However for $f_0\simeq 0.7-1$ that corresponds
to density typical of LHC also in this case $t_{BEC}\simeq 1\ \mathrm{fm /c}$. 
Considering that we are disregarding the $2\leftrightarrow 3$ processes that can 
accelerate significantly the dynamics \cite{Huang:2013lia,Xu:2008av}, it is conceivable that at highest LHC energy
one can enter into the region where even a pQCD dynamics can drive the system
into a BEC phase.

\section{Conclusions}

In this article, we have studied thermalization of a hot and dense homogeneous gluon gas in
 a box,
whose initial spectrum is of  glasma type with occupied states below the saturation scale, $Q_s$, and 
unpopulated states above $Q_s$. In order to study the evolution of the system from initial state towards
equilibration we have implemented a parton cascade code based on the solution of the kinetic equation,
by means of  a stochastic method  to compute the collision integral. 

For what concerns the numerical code, this is the first time that a parton cascade code studying a system of ultra-relativistic
particles
with a BE quantum kernel is presented. For this reason we have spent the first part 
of this paper to 
describe the necessary consistency checks of the outputs of the code; in particular
we checked that the fixed point of the quantum kinetic equation agrees with the 
analytical equilibrium distribution function which is expected for a given set of initial
particle and energy densities.

We have then focused on the evolution of the initial state
towards equilibrium. Our novelty, in comparison to previous studies, is that by using
the full kinetic equation we do not need to assume a small angle dominance of the cross section,
which justifies a Fokker-Planck approach. We go beyond
the small angle approximation of the kinetic equation by treating the Debye screening mass, $m_D$
in the cross section as a pure numerical parameter: the larger $m_D$ the more isotropic the cross section is.
Changing $m_D$ we have kept fixed the total cross section, in order to be sure that the
only change we introduce by the different $m_D$ is the change of the angular part of the differential
cross section. We have found that increasing $m_D$ lowers the thermalization time of about a factor 4 considering $m_D=1$ GeV with respect to the forward peaked case corresponding to $m_D=0.1$ GeV.

An important result of our study is the evolution of the system towards a BE condensate.
We have found, in agreement with previous studies, that if the initial density is large enough
the system evolves toward a BEC.
We have found that for values of $m_D\sim 1$ GeV which are relevant for heavy ion collisions the time
needed to form such a condensate could be as small as $t\sim0.2$ fm/c for densities comparable to those present in the final stage at LHC energy. 
Finally studying the pQCD case we observe that at phase space density $f_0\sim 1$ similar to that reached in the very early stage of LHC collisions $t_{\mathrm{BEC}}\lesssim 1$ fm/c.
Nonetheless, before giving quantitative estimates for heavy ion collisions we stress that our study
needs to be generalized to an expanding longitudinal geometry. Therefore we plan to implement
the longitudinal expansion in our quantum parton cascade code and to report on the effects
of the expansion.
\vspace{-0.85cm}
\section*{Appendix A. Collision rate}
In this appendix we derive an expression for the collision rate and for the collision probability.
Considering as a starting point the collision integral
\begin{eqnarray}
&&C^{out}[f]=\frac{1}{2E_{1}}\int \frac{d^3p_2}{ 2E_{2} (2\pi)^3}\frac{1}{\nu}
\int\frac{d^3p{^\prime}_1}{2E{^\prime}_1(2\pi)^3}\nonumber\\
&& \int\frac{d^3p{^\prime}_2}{ 2E{^\prime}_2(2\pi)^3}
f(p_1) f(p_2)(1+f(p{^\prime}_1))(1+f(p{^\prime}_2))\nonumber\\
&& \times|{\cal M}|^2\times(2\pi^4)\delta^4
(p_1 + p_2 - p_{1}^\prime - p_{2}^\prime)
\label{CollisionInt}
\end{eqnarray}
the collision rate $\Gamma$ can be expressed as
\begin{eqnarray}
\Gamma=\int \frac{d^3p_1}{(2\pi)^3}C^{out}[f]
\end{eqnarray}
thus
\begin{eqnarray}
&&\Gamma =\int \frac{d^3p_1}{ 2E_{1} (2\pi)^3} \int \frac{d^3p_2}{ 2E_{2} (2\pi)^3}\frac{1}{\nu}
\int\frac{d^3p{^\prime}_1}{2E{^\prime}_1(2\pi)^3}\nonumber\\
&& \int\frac{d^3p{^\prime}_2}{ 2E{^\prime}_2(2\pi)^3}
f(p_1) f(p_2)(1+f(p{^\prime}_1))(1+f(p{^\prime}_2))\nonumber\\
&& \times|{\cal M}|^2\times(2\pi^4)\delta^4
(p_1 + p_2 - p_{1}^\prime - p_{2}^\prime)
\end{eqnarray}
that can be written as
\begin{eqnarray}
&&\Gamma =\frac{1}{\nu}\int \frac{d^3p_1}{ (2\pi)^3} \int \frac{d^3p_2}{(2\pi)^3}f(p_1) f(p_2)\nonumber  \\
&& \times \int d\Omega \frac{d\sigma}{d\Omega}(1+f(p{^\prime}_1))(1+f(p{^\prime}_2)){v}_{rel}~.
\label{rate_cons1_text2}
\end{eqnarray}
We have evaluated numerically this integral following the same approach described in
\cite{PhysRevA.66.033606}.\\
From eq. (\ref{rate_cons1_text2}) one can get the expression 
for the collision probability 
used to evaluate the collision integral with the stochastic method.
In fact the the number of collision in a time step $\Delta t$ 
 in a volume $\Delta^3x$ for particle 
with momenta in the range $(p_1,p_1+ \Delta^3p_1)$ and $(p_2,p_2+\Delta^3p_2)$ 
can be written as 
\begin{eqnarray}
&&\Delta N_{coll}=\Delta^3x\Delta t \frac{\Delta^3p_1}{ (2\pi)^3} \frac{\Delta^3p_2}{(2\pi)^3}f(p_1) f(p_2)\nonumber  \\
&& \times \int d\Omega \frac{d\sigma}{d\Omega}(1+f(p{^\prime}_1))(1+f(p{^\prime}_2)){v}_{rel}~.
\label{rate_cons1_text3}
\end{eqnarray}
Expressing the distribution functions $f(p_1)$ and $f(p_2)$ 
as it has been done in \cite{Xu:2004mz}:
\begin{equation}
 f_i=\frac{\Delta N_i}{\frac{1}{(2\pi)^3}\Delta^3x\Delta^3p_i}
\end{equation}
and substituting in Eq. (\ref{rate_cons1_text3})
one gets 
\begin{eqnarray}
&&\Delta N_{coll}=\frac{\Delta N_1\Delta N_2\Delta^3t}{\Delta^3x} \nonumber \\
&& \times \int d\Omega \frac{d\sigma}{d\Omega}(1+f(p{^\prime}_1))(1+f(p{^\prime}_2)){v}_{rel}~.
\end{eqnarray}
Thus the number of collision for particles pairs $\Delta N_{coll}/\Delta N_1\Delta N_2$
 which is indeed
 the collision probability 
$P_{22}$, is given by
\begin{eqnarray}
P_{22}=\frac{\Delta t}{\Delta^3x} 
\int d\Omega \frac{d\sigma}{d\Omega}(1+f(p{^\prime}_1))(1+f(p{^\prime}_2)){v}_{rel}
\label{int_probability}
\end{eqnarray}
{\em Acknowledgements.}
The authors acknowledge discussions with J.Liao and N. Su.
V.~G., F. S. and D.P. acknowledge the ERC-STG funding under the QGPDyn grant.
V.G. thanks J.P. Blaizot for the kind hospitality at IPhT of Saclay that stimulated the present work.
\bibliography{iopart-num}{}
\bibliographystyle{h-physrev}
\end{document}